\documentclass[
reprint,
amsmath,amssymb,
aps,
prb,
floatfix]{revtex4-2}

\usepackage[english]{babel}
\usepackage{comment}
\usepackage{amsmath}
\usepackage{natbib}
\usepackage{graphicx}
\usepackage{dcolumn}
\usepackage{bm}
\usepackage{siunitx}
\usepackage{mathtools}
\usepackage{booktabs}
\usepackage{tabularx}
\usepackage{threeparttable}
\usepackage{array}
\usepackage[utf8]{inputenc}
\usepackage[T1]{fontenc}
\usepackage{xcolor}
\usepackage{url}
\usepackage{hyperref}
\hypersetup{colorlinks=true,allcolors=blue}
\graphicspath{{./}}

\newif\ifmarkup
\ifdefined\MARKUP\markuptrue\else\markupfalse\fi
\DeclareRobustCommand{\rev}[1]{\ifmarkup{\color{blue}#1}\else#1\fi}
\ifmarkup\hypersetup{hidelinks}\fi %

\begin{document}

\title{Gate-Controlled Spin Qubits in Confined Altermagnets}
\author{Hamed Vakili}
\affiliation{Department of Physics and Astronomy and Nebraska Center for Materials and Nanoscience, University of Nebraska, Lincoln, Nebraska 68588, USA}
\date{\today}

\begin{abstract}
We propose gate-defined spin qubits in electrostatically confined altermagnetic quantum dots, together with an electrical spin-drive channel that requires neither spin-orbit coupling (SOC) nor micromagnets or field gradients. Elliptical confinement projects the $d$-wave altermagnetic spin splitting onto the discrete dot spectrum, producing a low-energy qubit doublet whose splitting follows the gate geometry, with a microwave-range qubit gap and a meV-range leakage gap over the parameter ranges studied. Because the projected splitting tracks the confinement anisotropy, a phase-controlled resonant quadrupolar gate drive implements $X_{\pi/2}$ and $X_\pi$ rotations in time-dependent simulations. The drive matrix element is set by the non-relativistic altermagnetic hopping rather than by an SOC coefficient, with a static in-plane field of a few tenths of a tesla supplying the transverse mixing. \rev{The qubit has no first-order coupling to uniform electric-field noise, and its first-order quadrupolar charge-noise sensitivity vanishes at the circular point, where the drive is strongest.} Two-electron configuration-interaction calculations of a double dot yield an isolated logical two-qubit manifold with a nonzero nonlocal Pauli block, and resonant central-barrier modulation implements a perfectly entangling fSim gate. \rev{Within this model, confined altermagnets thus provide the ingredients for universal gate-based control from local electrostatic gates; we identify the material and device conditions that a realization must meet.}
\end{abstract}

\maketitle

\textit{Introduction}---Gate-defined spin qubits are a mature solid-state platform because electrostatic gates can define the confinement potential, tune tunnel couplings, and control readout conditions in semiconductor nanostructures~\cite{Loss1998,Hanson2007,Vandersypen2017,Burkard2023}. After single-shot readout, exchange control, and electrically driven spin resonance were established in GaAs quantum dots (QD)~\cite{Elzerman2004,Petta2005,Koppens2006,Nowack2007}, silicon, donor, and germanium devices added long-coherence hosts, fault-tolerance-threshold two-qubit logic, multi-qubit and hole-spin processors, and foundry-compatible unit cells~\cite{Maune2012,Pla2012,Veldhorst2014,Yoneda2018,Zajac2018,Watson2018,Huang2019,Hendrickx2020Ge,Hendrickx2020Hole,Hendrickx2021GeProcessor,Noiri2022,Xue2022,Mills2022,Philips2022,Scappucci2021,Steinacker2025,Madzik2025}. In essentially all electrically driven implementations, the gate voltage couples to the spin through one of three channels: relativistic spin-orbit coupling (SOC) via electric-dipole spin resonance (EDSR), an engineered field gradient from a micromagnet, or electrical modulation of an SOC-derived $g$-tensor~\cite{Kato2003,Crippa2018}. \rev{These approaches are highly developed, but their drive strength is set by relativistic energy scales or by fabricated gradients, micromagnets add stray fields and fabrication overhead, and the electric-dipole coupling that enables driving also exposes the spin to charge noise. This motivates the search for an additional electrical spin-drive channel that is available in materials with no net magnetic moment and operates without SOC.}

Altermagnets provide such a route. They are collinear compensated magnets whose crystal symmetries permit momentum-dependent spin splitting~\cite{Smejkal2022PRX,Smejkal2022AM,Mazin2022}. Altermagnetic spin splitting has been reported experimentally in several systems~\cite{Krempasky2024,Reimers2024}, and related transport and hybrid-device consequences have been studied theoretically~\cite{Ouassou2023,Beenakker2023,Giil2024,PhysRevLett.133.226002,AM_diode,Vakili2025,AM_Thermomagnonic,kqy8-myz1,zn7r-k1xd,Geometry_AM_Soori,SC_AM_Vakili,Fu2025,SciPostPhys.20.2.059,Tjernshaugen2026}. In a gate-defined dot, the electronic states are discrete bound states rather than Bloch states labeled by crystal momentum. The central question is how the $d$-wave spin splitting survives projection onto the bound-state spectrum, and whether the resulting states can serve as qubit logical states.

\begin{figure}[!htbp]
    \centering
    \includegraphics[width=\columnwidth]{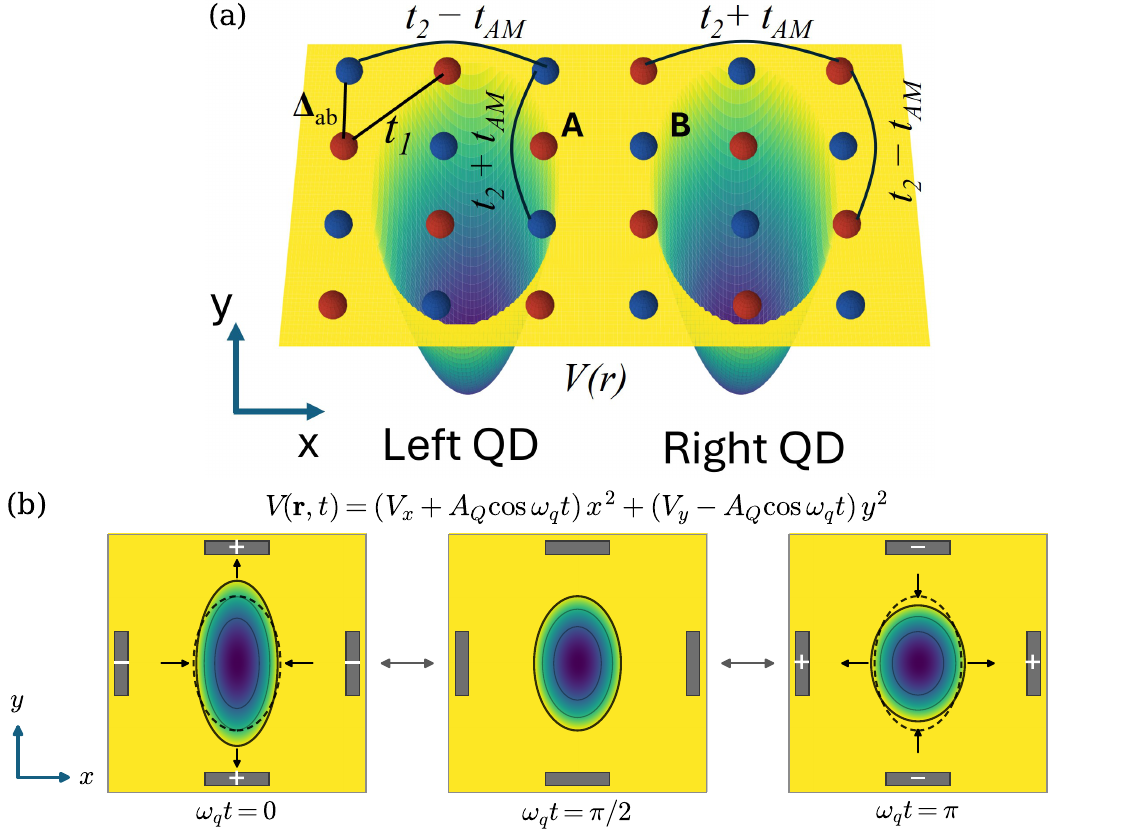}
    \caption{(a) Schematic device geometry. A two-dimensional $d$-wave altermagnetic layer is confined by electrostatic gates. For the two-qubit device, two locally anisotropic gates define left and right dots, while a central barrier gate sets the static interdot coupling and supplies resonant barrier modulation. \rev{(b) Quadrupolar single-qubit gating, top view: RF voltages of opposite sign on the $x$ and $y$ gate pairs deform the dot about its elliptical bias shape (dashed); modulation exaggerated.}}
    \label{fig:device}
\end{figure}

Here we address this question with an atomistic real-space implementation of the minimal tight-binding model for a 2D $d$-wave altermagnet~\cite{Minimal_models}. \rev{Its continuum limit is the generic two-band $d$-wave form~\cite{Smejkal2022PRX,Minimal_models}, and a lattice fitted to first-principles coefficients gives the same doublet and drive~\cite{Note},} so the effect is not an artifact of one parametrization. The device concept is sketched in Fig.~\ref{fig:device}: an elliptical gate defines a single altermagnetic dot for one-qubit operation, while a pair of such gates plus a tunable barrier defines a double dot for two-qubit operation. Elliptical confinement idealizes noncircular gate-defined dots, where gates routinely reshape single-particle spectra and anisotropic response~\cite{Kouwenhoven1997,Kyriakidis2002,Allen2012,Prabhakar2009,Leon2020,Hendrickx2024SweetSpot}. Without SOC, in-plane parity forbids any linear electric-dipole spin matrix element in a centered dot, so conventional EDSR is unavailable by symmetry; the leading symmetry-allowed electrical channel is instead the even-parity quadrupolar shape mode, which modulates the confinement-projected altermagnetic field itself. The scheme is thus a single-spin, non-relativistic analog of $g$-tensor modulation resonance~\cite{Kato2003,Crippa2018}: the modulated splitting is supplied by the exchange-set altermagnetic hopping $t_{AM}$ rather than by SOC, so the control matrix element scales with a band energy rather than with a relativistic coefficient.

\textit{Hamiltonian and QD setup}---The simulations use a finite square atomistic lattice of linear size $L$ with open boundaries; $a$ is the nearest-neighbor A--B site spacing in the checkerboard lattice, with sublattice label $\eta_i=+1~(-1)$ on A~(B) sites.
The operator $c_{is}^{\dagger}$ creates an electron with spin $s=\uparrow,\downarrow$ on atom $i$, and $c_i=(c_{i\uparrow},c_{i\downarrow})^T$. The one-electron Hamiltonian is
$
H_{\mathrm{1e}}=
H_{\mathrm{hop}}+H_{\mathrm{ex}}+H_Z+H_V .
\label{eq:H1e}
$
The spin-independent part is
\begin{align}
H_{\mathrm{hop}}={}&
\Delta_{ab}\sum_{\langle ij\rangle,s}
\left(c_{is}^{\dagger}c_{js}+\mathrm{H.c.}\right)
+t_1\sum_{\langle\langle ij\rangle\rangle,s}
\left(c_{is}^{\dagger}c_{js}+\mathrm{H.c.}\right)
\nonumber\\
&+\sum_{i,s}
\left[
(t_2+\eta_i t_{AM})c_{is}^{\dagger}c_{i+2\hat{x},s}
+\mathrm{H.c.}
\right]
\nonumber\\
&+\sum_{i,s}
\left[
(t_2-\eta_i t_{AM})c_{is}^{\dagger}c_{i+2\hat{y},s}
+\mathrm{H.c.}
\right]
-\mu\sum_{i,s}n_{is}.
\label{eq:Hhop_atomistic}
\end{align}
Here $\langle ij\rangle$ denotes cardinal nearest-neighbor bonds between opposite sublattices, $\langle\langle ij\rangle\rangle$ diagonal same-sublattice bonds, and the $t_2,~t_{AM}$ hoppings along $2\hat{x}$ and $2\hat{y}$ are also same-sublattice. The sublattice-odd $\eta_i t_{AM}$ term is the atomistic origin of the $d$-wave spin splitting. 

The staggered exchange field is local on the atomistic lattice,
$
H_{\mathrm{ex}}=
\frac{J}{2}
\sum_i
\eta_i\,
c_i^\dagger(\mathbf{L}\cdot\bm{\sigma})c_i ,
\label{eq:Hex}
$
where $\mathbf{L}$ is the altermagnetic spin axis, taken along $\hat{z}$ unless stated otherwise. The in-plane magnetic field is included, with $g=2$, through
$
H_Z=
\frac{g\mu_B B_x}{2}
\sum_i
c_i^\dagger\sigma_x c_i .
\label{eq:HZ}
$
No spin-orbit coupling is included in the main-text Hamiltonian, which leaves $B_x$ as the only spin-mixing term. \rev{For a two-dimensional layer the in-plane field enters only through $H_Z$, and $\mathbf{L}$ is kept rigid: a field perpendicular to $\mathbf{L}$ causes no spin flop, and the canting it induces only renormalizes the transverse field (Supplemental Material~\cite{Note}). All single-dot calculations use $L=40$~nm, for which the spectra are converged with respect to the box size~\cite{Note}.}

\textit{Single-dot confinement.}---For a single QD we use
\begin{equation}
\begin{aligned}
H_V={}&
\sum_{i,s}
V(\mathbf{r}_i)c_{is}^\dagger c_{is},~
V(\mathbf{r})=
V_xx^2+V_yy^2,
\end{aligned}
\label{eq:Vsingle}
\end{equation}
where $V_x$ and $V_y$ are the electrostatic potential curvatures along the two in-plane directions. The ellipticity parameter is
$
\delta\equiv
\frac{V_y-V_x}{V_y+V_x}.
\label{eq:delta}
$
In the $\delta$ scans below, the mean curvature $\bar V=(V_x+V_y)/2$ is held fixed unless stated otherwise.

\begin{figure}[!htbp]
    \centering
    \includegraphics[width=\columnwidth]{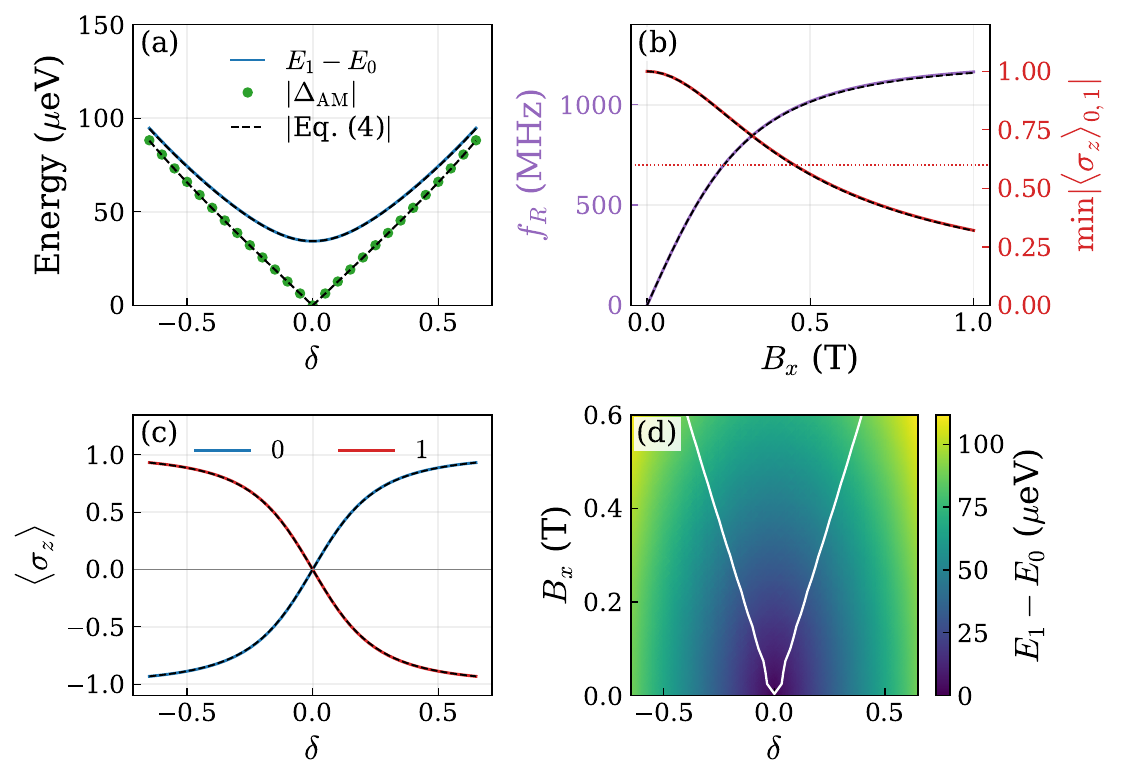}
    \caption{\rev{(a) Qubit splitting $E_1-E_0$ at $B_x=\SI{0.3}{\tesla}$ (solid) and field-free splitting $|\Delta_{\rm AM}|$ (circles) versus ellipticity $\delta$; dashed lines are the magnitude of Eq.~(\ref{eq:dwave_projection}) and the two-level expression for $E_1-E_0$. (b) Dependence on $B_x$ at $\delta=0.3$: rotating-wave Rabi frequency $f_R=A_Q|\langle1|Q|0\rangle|/h$ for $A_Q=10^{-5}~\mathrm{eV\,nm^{-2}}$ and logical-state polarization $\min|\langle\sigma_z\rangle_{0,1}|$; dashed lines are the two-level predictions $f_R\propto\sin\theta$ and $\cos\theta$, and the dotted line marks 0.6.} (c) Spin polarizations of the two lowest states at $B_x=\SI{0.3}{\tesla}$; dashed lines, $\pm\cos\theta$. (d) Color map of $E_1-E_0$ versus $\delta$ and $B_x$; the white contour $\min|\langle\sigma_z\rangle_{0,1}|=0.6$ separates the nearly circular region (inside) from the region where the two lowest states have opposite spin polarization of magnitude at least 0.6 (outside). Parameters: $a=\SI{0.5}{\nano\meter}$, $L=\SI{40}{\nano\meter}$, $t_1=-\SI{150}{\milli\electronvolt}$, $t_2=0$, $t_{AM}=\SI{20}{\milli\electronvolt}$, $\Delta_{ab}=\SI{8}{\milli\electronvolt}$, $J=\SI{10}{\milli\electronvolt}$, $\mu=0$, no SOC, fixed mean curvature $\bar V=0.13~\mathrm{meV\,nm^{-2}}$. At the representative point $\delta=0.3$, $B_x=\SI{0.3}{\tesla}$, $(V_x,V_y)=(0.091,0.169)~\mathrm{meV\,nm^{-2}}$, $E_1-E_0=\SI{51.7}{\micro\electronvolt}$ ($12.5$~GHz), $\Delta_{\rm AM}=\SI{38.7}{\micro\electronvolt}$, $\langle\sigma_z\rangle_{0,1}=\pm0.75$, and the leakage gap is $E_2-E_0=\SI{5.2}{\milli\electronvolt}$.}
    \label{fig:static}
\end{figure}

\rev{The connection between the atomistic $t_{AM}$ term and the dot spectrum follows from the continuum limit of the lowest band. Near $\Gamma$, the two-sublattice Bloch Hamiltonian of Eq.~(\ref{eq:Hhop_atomistic}) gives spin-dependent anisotropic masses,
\begin{equation}
E_\sigma(\mathbf{k})\simeq E_{\min}+(A+\sigma\lambda)k_x^2+(A-\sigma\lambda)k_y^2,
\label{eq:continuum_band}
\end{equation}
with $A=a^2(4\Delta_{ab}^2/R_0-2t_1)$, $\lambda=2Jt_{AM}a^2/R_0$, $R_0=[(J/2)^2+16\Delta_{ab}^2]^{1/2}$, $t_2=0$, and $\sigma=\pm1$ for $s=\uparrow,\downarrow$: the sublattice-odd $t_{AM}$ hopping acts on the lowest band through its exchange-induced sublattice polarization $J/2R_0$. For $V=V_xx^2+V_yy^2$ the dot levels then follow exactly, $E_{n\sigma}=E_{\min}+2\sqrt{(A+\sigma\lambda)V_x}\,(n_x+\tfrac12)+2\sqrt{(A-\sigma\lambda)V_y}\,(n_y+\tfrac12)$, and the field-free splitting of the lowest doublet is
\begin{align}
\Delta_{\rm AM}&=\big(\sqrt{A+\lambda}-\sqrt{A-\lambda}\big)\big(\sqrt{V_y}-\sqrt{V_x}\big)
\nonumber\\
&\simeq\frac{\lambda}{\sqrt{A}}\sqrt{\bar V}\,\delta\Big(1+\frac{\delta^2}{8}\Big).
\label{eq:dwave_projection}
\end{align}
Equation~(\ref{eq:dwave_projection}) makes the projection argument quantitative: the splitting is linear in $\delta$ up to an $O(\delta^3)$ correction, vanishes for a circular dot, reverses sign with $\delta$, and is linear in $t_{AM}$ through $\lambda$ for $\lambda\ll A$; an excited orbital is split in proportion to $(n_y+\tfrac12)\sqrt{V_y}-(n_x+\tfrac12)\sqrt{V_x}$. With the field, the doublet is the two-level system $H_q=-\tfrac12\Delta_{\rm AM}\sigma_z+\tfrac12g_\perp\mu_BB_x\sigma_x$, so $E_1-E_0=[\Delta_{\rm AM}^2+(g_\perp\mu_BB_x)^2]^{1/2}$ and $\langle\sigma_z\rangle_{0,1}=\pm\cos\theta$ with $\tan\theta=g_\perp\mu_BB_x/\Delta_{\rm AM}$, where $g_\perp\simeq0.99\,g$ contains the sublattice and orbital overlaps of the two spin states~\cite{Note}. For Fig.~\ref{fig:static}, $A=77.0$~meV\,nm$^2$ (band mass $0.49\,m_e$), $\lambda=3.09$~meV\,nm$^2$, and $2R_0=65$~meV; these expressions reproduce the atomistic gaps, polarizations, and quadrupolar matrix elements [dashed lines in Fig.~\ref{fig:static}].}

\textit{Single-qubit regime.}---Fig.~\ref{fig:static} shows that the atomistic model supports an isolated low-energy doublet with opposite spin polarization: $E_1-E_0$ stays in the microwave range over the plotted ellipticity interval while the leakage gap remains far larger.

Reversing $\delta$ reverses $\Delta_{\rm AM}$, so the two lowest states exchange spin-sublattice character across the circular point [Fig.~\ref{fig:static}(c)].

\rev{The static field sets the operating regime [Fig.~\ref{fig:static}(b)]. Because $Q$ is spin independent, $|\langle1|Q|0\rangle|=\tfrac12|Q_\uparrow-Q_\downarrow|\sin\theta$, with $Q_\sigma$ the quadrupole moment of the spin-$\sigma$ orbital: the drive vanishes at $B_x=0$, grows linearly at small field, and saturates for $g_\perp\mu_BB_x\gg\Delta_{\rm AM}$, while the polarization decreases as $\cos\theta$. Increasing $B_x$ thus trades spin polarization for drive efficiency.} Fig.~\ref{fig:static}(d) is an operating map: the white contour ($\cos\theta=0.6$, i.e., $g_\perp\mu_BB_x=\tfrac43|\Delta_{\rm AM}|$) separates the nearly circular regime from the region where the doublet has well-defined spin-sublattice character.

The Supplemental Material~\cite{Note} reports one- and two-parameter sweeps over $J$, $t_{AM}$, $\Delta_{ab}$, and interfacial Rashba coupling: the microwave qubit gap, meV-scale leakage gap, and finite quadrupolar matrix element persist over a broad parameter range, with trends that follow $\lambda\propto Jt_{AM}/R_0$.

For one-qubit control we use a resonant modulation of the quadrupolar confinement anisotropy about a fixed operating point, weak on the scale of the qubit subspace,
\begin{equation}
H(t)=H_{\mathrm{1e}}+
A_Q\cos(\omega_q t+\phi)\,Q,
~~
Q=x^2-y^2 .
\label{eq:resonant_quad_drive}
\end{equation}
Here $H_{\mathrm{1e}}$ is the static Hamiltonian at the operating point, $\omega_q=(E_1-E_0)/\hbar$, and $A_Q$ (units of energy per area) absorbs the gate-voltage calibration. Radio-frequency (RF) voltages on the $x$ and $y$ gate pairs, in anti\-phase, generate this even-parity shape mode [Fig.~\ref{fig:device}(b)]; related quadrupole modes appear in charge-quadrupole and circuit-coupled dot devices~\cite{Friesen2017,Koski2020,Kratochwil2021}. In the low-energy description, the drive changes the projected $d$-wave field transverse to the dressed qubit axis; pulse area sets the rotation angle and carrier phase the rotation axis~\cite{Golovach2006,Tokura2006,PioroLadriere2008,McKay2017,Noiri2022,Xue2022,Philips2022,Steinacker2025}. Virtual $Z_\theta$ operations, including the non-Clifford $T=Z_{\pi/4}$, are implemented as phase-frame updates of subsequent pulses~\cite{McKay2017}. Time evolution is propagated by stepwise midpoint matrix exponentials (time step $0.5$~ps) in the basis of the lowest 64 eigenstates at the operating point; halving the time step or enlarging the basis to 96 states changes the gate fidelities by less than $10^{-6}$~\cite{Note}.

\begin{figure}[!htbp]
    \centering
    \includegraphics[width=\columnwidth]{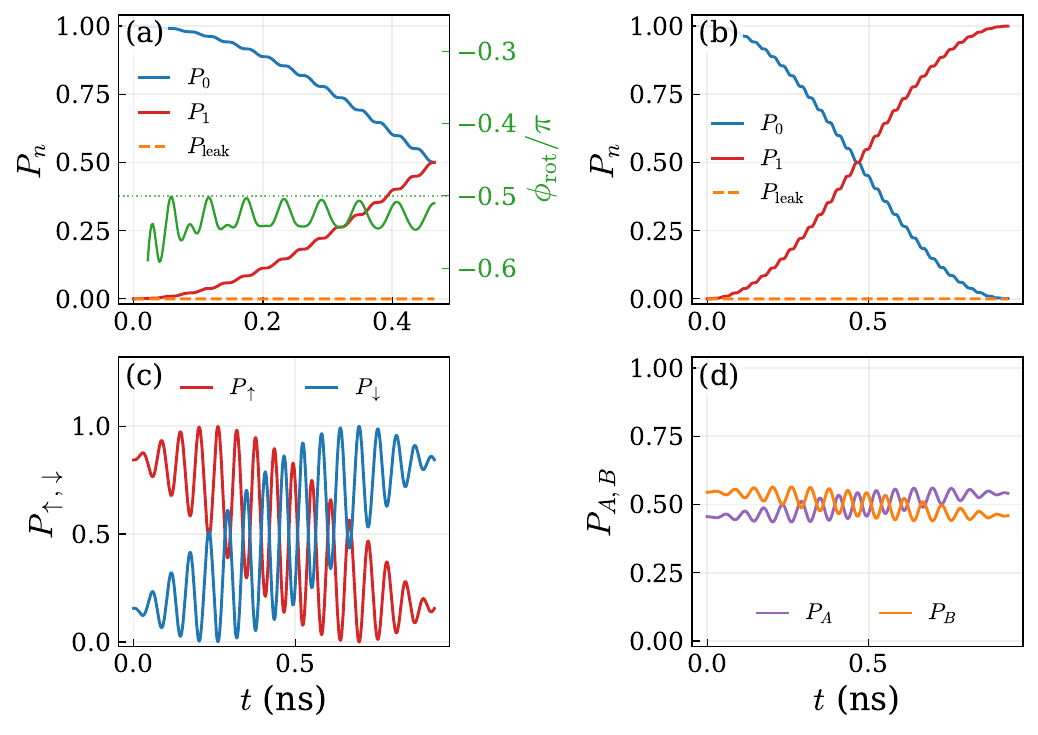}
    \caption{Single-qubit operation by a phase-controlled resonant quadrupolar drive. (a) $X_{\pi/2}$ pulse from $|0\rangle$, showing $P_0$, $P_1$, $P_{\mathrm{leak}}$, and the rotating-frame relative phase $\phi_{\mathrm{rot}}=\arg(c_1/c_0)+\omega_q t$ relative to the pulse axis; the dotted line marks $-\pi/2$. (b) $X_\pi$ pulse producing a $|0\rangle\rightarrow |1\rangle$ transition. (c) Spin weights $P_{\uparrow}$ and $P_{\downarrow}$ during the $X_\pi$ pulse. (d) Sublattice weights $P_A$ and $P_B$ during the $X_\pi$ pulse. Parameters: $a=\SI{0.5}{\nano\meter}$, $L=\SI{40}{\nano\meter}$, $t_1=-\SI{150}{\milli\electronvolt}$, $t_2=0$, $t_{AM}=\SI{15}{\milli\electronvolt}$, $\Delta_{ab}=\SI{12}{\milli\electronvolt}$, $J=\SI{12}{\milli\electronvolt}$, $\mu=0$, $B_x=\SI{0.45}{\tesla}$, no SOC, curvatures $(V_x,V_y)=(0.18,0.43)~\mathrm{meV\,nm^{-2}}$ ($\delta=0.41$); $E_1-E_0=\SI{71.1}{\micro\electronvolt}$, $f_q=\SI{17.2}{\giga\hertz}$, $\langle\sigma_z\rangle_{0,1}=\pm0.69$, and $E_2-E_0=\SI{7.4}{\milli\electronvolt}$. $A_Q=1.49\times10^{-5}\,\mathrm{eV\,nm^{-2}}$ ($8\%$ of $V_x$), giving $|\langle 1|A_Q Q|0\rangle|=\SI{2.2}{\micro\electronvolt}$. The $X_{\pi/2}$ and $X_\pi$ pulses last 8 and 16 carrier cycles ($0.47$ and $0.93$~ns) and start at a node of the drive ($\phi=\pi/2$). The average gate fidelities are $\bar F=0.9997$ ($X_{\pi/2}$) and $0.9993$ ($X_\pi$); the final $X_\pi$ transfer gives $P_1=0.999$, $P_0=0.001$, and $P_{\mathrm{leak}}\approx10^{-6}$.}
    \label{fig:gate}
\end{figure}

In Fig.~\ref{fig:gate}, $P_n(t)=|\langle\psi_n|\Psi(t)\rangle|^2$ with $\psi_n$ the low-energy eigenstates at the initial operating point and $P_{\mathrm{leak}}=1-P_0-P_1$; the spin and sublattice weights sum $|\Psi|^2$ over spin components and over sites with $\eta_i=\pm1$, respectively.

In Fig.~\ref{fig:gate}, the $X_{\pi/2}$ pulse produces nearly equal $P_0$ and $P_1$ while the rotating-frame phase approaches the expected $-\pi/2$ value. Doubling the pulse area gives a nearly complete $X_\pi$ transition with negligible leakage; the residual infidelity is a coherent error from the counter-rotating term and off-resonant orbital levels that decreases with pulse length, and both depend on the switch-on phase of the rectangular pulse~\cite{Note}. Panels (c,d) show the operation is not a charge excitation: the gate modulates the anisotropy, the altermagnetic term converts it into a spin-sublattice bias, and $B_x$ supplies the transverse mixing needed for coherent rotations.

Because the transverse coupling derives from $t_{AM}$, an exchange-scale band parameter, the drive remains available in light-element altermagnetic semiconductors whose weak SOC would starve conventional EDSR. Conversely, when interfacial Rashba coupling is present, the same dot additionally supports zero-applied-field EDSR through the parity-odd dipole channel (Supplemental Material~\cite{Note}), so both electrical control channels coexist in one device and are distinguished by the parity of the drive.

\rev{\textit{Noise sensitivity.}---The parity argument that forbids a dipolar spin drive also removes the leading dipolar noise channel: a uniform in-plane field only displaces a harmonic dot, so $E_1-E_0$ has no first-order dependence on it. Shape fluctuations couple through the drive operator itself, whose longitudinal and transverse parts in the qubit basis have the fixed ratio $|Q_{11}-Q_{00}|/2|Q_{01}|=\cot\theta$; the mixing that enables driving also reduces this dephasing. At the circular point the first-order shape sensitivity vanishes by symmetry while the drive is strongest, a quadrupolar sweet spot analogous to hole-spin sweet spots~\cite{Hendrickx2024SweetSpot}. For rms electrochemical-potential noise $\sigma_\mu\approx1.3$--$4.3~\mu$eV, typical of Si/SiGe~\cite{PaqueletWuetz2023,Note}, an ensemble of remote charge fluctuators gives a charge-noise-limited $T_2^*\approx0.06$--$0.3$~s at this point, against $0.6$--$2~\mu$s at the point of Fig.~\ref{fig:gate}, where $\partial f_q/\partial\delta=21$~GHz. The protection is partial: it costs spin polarization, and field noise still gives $T_2^*\approx8~\mu$s at $\sigma_B=1~\mu$T~\cite{Note}. Quasi-static noise ensembles of the Fig.~\ref{fig:gate} gates~\cite{Note} give noise-induced infidelities below $10^{-4}$ up to $\sigma_\mu\approx60~\mu$eV, field noise of $\approx0.2$~mT, and relative exchange or N\'eel-vector fluctuations of order $10^{-3}$, so at that point $T_2^*/t_{\rm gate}\sim10^3$, rather than gate error, sets the budget.}

\begin{figure}[!htbp]
    \centering
    \includegraphics[width=0.96\columnwidth]{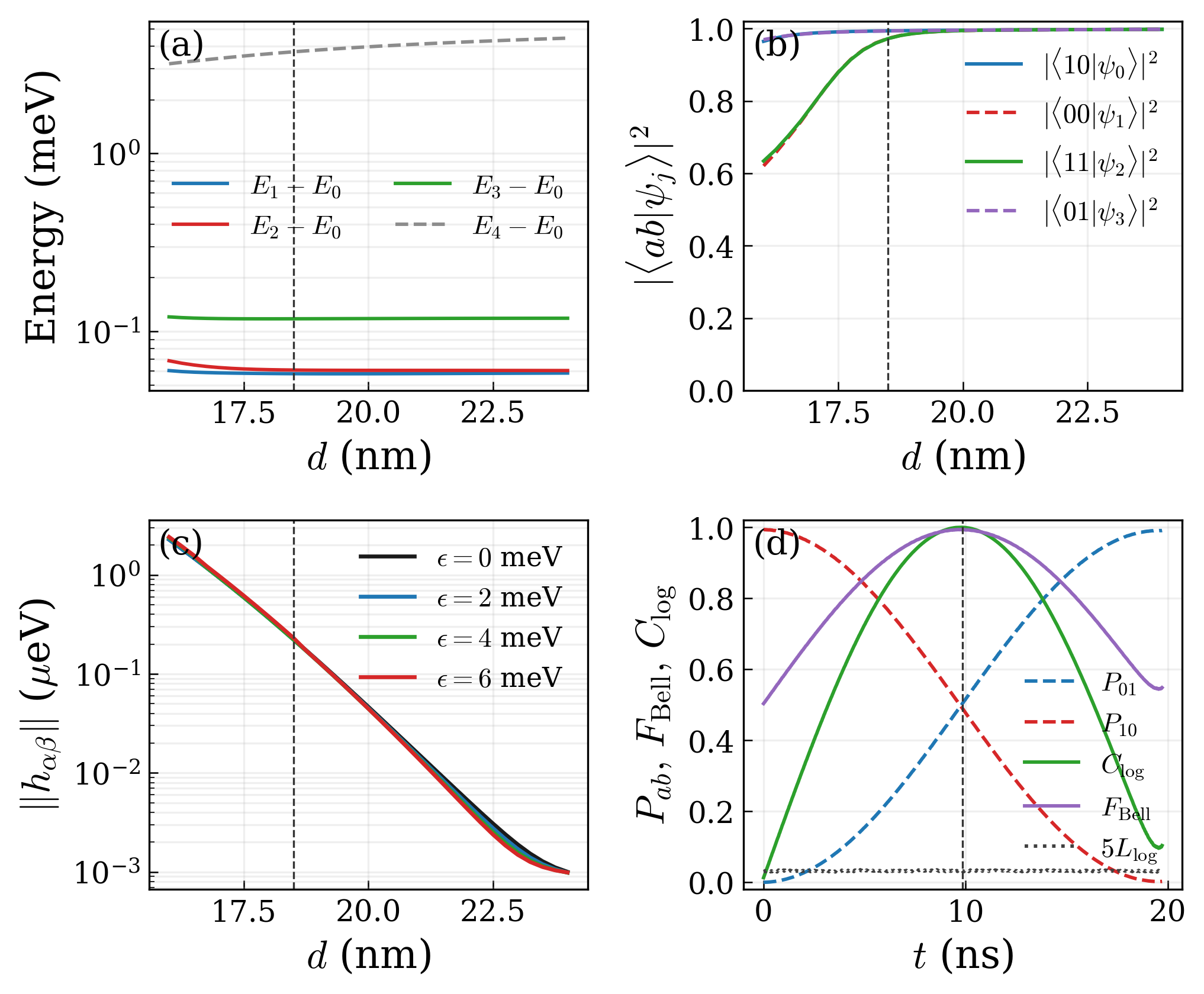}
    \caption{Two-electron double-dot calculation using the atomistic Hamiltonian and a truncated CI basis.
    (a) Four-state CI manifold versus dot separation $d$, shown as energy gaps relative to $E_0$ together with the first leakage level $E_4-E_0$.
    (b) Isolated QD product overlaps $|\langle ab|\psi_j\rangle|^2$ for the four lowest dressed CI eigenstates.
    (c) Frobenius norm $\|h_{\alpha\beta}\|$ of the $3\times3$ nonlocal Pauli block in Eq.~(\ref{eq:nonlocal_pauli_block}) for detunings $\epsilon=0,2,4,6~\mathrm{meV}$.
    (d) Localized-basis populations and entanglement diagnostics during weak resonant barrier modulation at $d=\SI{18.5}{\nano\meter}$; $5L_{\rm log}$ is plotted, and the dashed vertical line marks the maximum Bell fidelity.
    Parameters: material parameters as in Fig.~\ref{fig:gate}, $L=\SI{60}{\nano\meter}$, $B_x=\SI{0.3}{\tesla}$; left-dot curvatures $(V_x^L,V_y^L)=(0.185,0.434)~\mathrm{meV\,nm^{-2}}$, right-dot $(V_x^R,V_y^R)=(0.434,0.185)~\mathrm{meV\,nm^{-2}}$; $d\in[\SI{16}{\nano\meter},\SI{24}{\nano\meter}]$ in panels (a)--(c), static detuning $\epsilon=\SI{4}{\milli\electronvolt}$ except where varied in panel (c), $V_s=\SI{0.2}{\milli\electronvolt}$, $(w_x,w_y)=(\SI{8}{\nano\meter},\SI{12}{\nano\meter})$, $V_b^{(0)}=0$, $(\epsilon_r,d_{\rm soft},\ell_s)=(24,\SI{2.5}{\nano\meter},\SI{40}{\nano\meter})$, $M=32$.
    The barrier drive has amplitude $\SI{0.5}{\milli\electronvolt}$, frequency $\omega_{10,01}$, and duration $\approx \SI{20}{\nano\second}$ over $561$ carrier cycles.
    At the Bell point $t=\SI{9.83}{\nano\second}$, $P_{01}=0.499$, $P_{10}=0.496$, $C_{\rm log}=0.9998$, $F_{\rm Bell}=0.996$, and $L_{\rm log}=4.1\times10^{-3}$.}
    \label{fig:twoqubit}
\end{figure}

\textit{Two-qubit configuration.}---For the two-electron configuration-interaction (CI) calculation, we use the same atomistic real-space tight-binding model in a double-dot potential. The local left and right wells are
\begin{equation}
\begin{aligned}
V_L(x,y)={}&V_x^L(x+d/2)^2+V_y^L y^2+\epsilon/2,\\
V_R(x,y)={}&V_x^R(x-d/2)^2+V_y^R y^2-\epsilon/2,
\end{aligned}
\label{eq:VLVR}
\end{equation}
where $d$ is the dot separation, $\epsilon$ is the detuning, and $V_{x,y}^{L,R}$ set the local confinement curvatures. The double-well floor is a smooth minimum of the two parabolas,
\begin{equation}
V_{\mathrm{dot}}(x,y)
=
-V_s\ln\!\left[
e^{-V_L(x,y)/V_s}+e^{-V_R(x,y)/V_s}
\right],
\label{eq:Vdot_smooth}
\end{equation}
where $V_s$ smooths the interdot saddle, and the full double-dot potential adds a central Gaussian barrier,
\begin{equation}
V_{\mathrm{DD}}(x,y)
=
V_{\mathrm{dot}}(x,y)
+V_b\exp[-(x/w_x)^2-(y/w_y)^2],
\label{eq:VDD}
\end{equation}
where $V_b$ is the central-barrier amplitude and $w_x,w_y$ are the barrier widths. In the atomistic Hamiltonian, $V_{\rm DD}$ enters as the scalar gate potential $H_V^{\rm DD}(V_b)=\sum_{i,s}V_{\rm DD}(\mathbf{r}_i;V_b)c_{is}^\dagger c_{is}$, with $H_{\rm 1e}^{\rm DD}(V_b)=H_{\rm hop}+H_{\rm ex}+H_Z+H_V^{\rm DD}(V_b)$. For the driven two-qubit calculation, the local wells and detuning are held fixed and only the central barrier is modulated,
\begin{equation}
V_b(t)=V_b^{(0)}+A_b\sin(\omega_{ab,cd}t),
~~
\omega_{ab,cd}=\frac{E_{j_{cd}}-E_{j_{ab}}}{\hbar},
\label{eq:barrier_drive}
\end{equation}
where $ab$ ($cd$) is the starting (target) localized product state and $j_{ab}$, $j_{cd}$ are the corresponding dressed CI eigenstate indices at $V_b^{(0)}$. For the Fig.~\ref{fig:twoqubit}(d) transfer, $ab=10$, $cd=01$, $j_{10}=0$, and $j_{01}=3$.

We use a truncated CI calculation, standard for few-electron quantum dots and exchange-coupled spin-qubit devices~\cite{Reimann2002,Rontani2006,Burkard1999,HuDasSarma2000,Nielsen2012,Hetenyi2020,Joecker2021,TariqHu2022,Geyer2024,Rodriguez2025}. For each static double-dot geometry we diagonalize $H_{\rm 1e}^{\rm DD}(V_b)$, $H_{\rm 1e}^{\rm DD}\psi_i=\varepsilon_i\psi_i$ (with $\varepsilon_i$ distinct from the detuning $\epsilon$), and retain the lowest $M=32$ eigenstates. The operators $d_i^\dagger=\sum_{p,s}\psi_{is}(p)c_{ps}^\dagger$ and $d_i=\sum_{p,s}\psi_{is}^\ast(p)c_{ps}$ create and annihilate electrons in the retained eigenstates, and the two-electron basis consists of antisymmetrized Slater determinants built from them. In this basis,
\begin{equation}
H_{\mathrm{CI}}(V_b)
=
\sum_i \varepsilon_i(V_b) d_i^\dagger d_i
+\frac{1}{2}
\sum_{ijkl}
V_{ij;kl}(V_b)
d_i^\dagger d_j^\dagger d_l d_k ,
\label{eq:HCI}
\end{equation}
where the Coulomb matrix elements $V_{ij;kl}=\int d^2r\,d^2r'\,\rho_{ik}(\mathbf{r})V_C(\mathbf{r}-\mathbf{r}')\rho_{jl}(\mathbf{r}')$ retain the spinor structure through the component-summed pair densities $\rho_{ik}(\mathbf{r}_p)=\sum_{s}\psi_{is}^\ast(p)\psi_{ks}(p)$ on atomistic sites, and $V_C(\mathbf{r})=e^2e^{-r/\ell_s}/[4\pi\epsilon_0\epsilon_r(r^2+d_{\rm soft}^2)^{1/2}]$ is a softened screened two-dimensional interaction with softening length $d_{\rm soft}$ and screening length $\ell_s$.

To diagnose two-qubit behavior, we build fixed localized reference states from isolated left and right dots (same atomistic parameters, neighboring dot removed), taking the lowest negative- and positive-$\sigma_z$ states as the local $0$ and $1$, and antisymmetrize the two-qubit product states, $|ab\rangle_{\mathrm{loc}}=[\phi_{L,a}(1)\phi_{R,b}(2)-\phi_{R,b}(1)\phi_{L,a}(2)]/\sqrt{2}$. If $W$ is the matrix whose columns are the four orthonormalized localized states, ordered as $|00\rangle_{\rm loc}$, $|01\rangle_{\rm loc}$, $|10\rangle_{\rm loc}$, and $|11\rangle_{\rm loc}$, then $H_{\rm log}=W^\dagger H_{\rm CI}W=\sum_{\mu,\nu\in\{I,x,y,z\}}h_{\mu\nu}\,\sigma_L^\mu\otimes\sigma_R^\nu$ with $h_{\mu\nu}=\tfrac14\mathrm{Tr}[(\sigma_L^\mu\otimes\sigma_R^\nu)H_{\rm log}]$. Terms with $\mu=I$ or $\nu=I$ are single-qubit fields and offsets, while the genuine two-qubit part is
\begin{equation}
H_{\rm nonlocal}
=
\sum_{\alpha,\beta=x,y,z}
h_{\alpha\beta}\,
\sigma_L^\alpha\otimes\sigma_R^\beta ,
\label{eq:nonlocal_pauli_block}
\end{equation}
whose first-order static strength we quantify by $\|h_{\alpha\beta}\|=[\sum_{\alpha,\beta=x,y,z}|h_{\alpha\beta}|^2]^{1/2}$. For the dynamical panel in Fig.~\ref{fig:twoqubit}, the retained-CI state is projected onto the same fixed basis: writing $c_{ab}=\langle ab|\Psi(t)\rangle$, $W_{\rm log}=\sum_{ab}|c_{ab}|^2$, and $\rho(t)=|\Psi(t)\rangle\langle\Psi(t)|$, the diagnostics are $P_{ab}=|c_{ab}|^2$, $L_{\rm log}=1-W_{\rm log}$, $C_{\rm log}=2|c_{00}c_{11}-c_{01}c_{10}|/W_{\rm log}$, and $F_{\rm Bell}=\max_{\phi}\langle\Phi_\phi|\rho(t)|\Phi_\phi\rangle$ with $|\Phi_\phi\rangle=(|01\rangle_{\rm loc}+e^{i\phi}|10\rangle_{\rm loc})/\sqrt{2}$. The time-dependent Hamiltonian is built by projecting the barrier-drive operator in Eq.~(\ref{eq:barrier_drive}) into the fixed dressed-CI basis at the operating point and is evolved with the same stepwise midpoint exponential propagator (time step 2~ps; halving it changes the two-qubit process fidelity by $<10^{-5}$).

Fig.~\ref{fig:twoqubit}(a) shows the four-state CI manifold separated from the first leakage level across the separation range. At $d=\SI{18.5}{\nano\meter}$, $\epsilon=\SI{4}{\milli\electronvolt}$, and $V_b^{(0)}=0$, the four lowest dressed CI states are assigned to $|10\rangle_{\rm loc}$, $|00\rangle_{\rm loc}$, $|11\rangle_{\rm loc}$, and $|01\rangle_{\rm loc}$ with minimum product overlap $0.97$ and leakage gap $E_4-E_3=3.6~\mathrm{meV}$; this assignment is stable with $M$~\cite{Note}.

The Pauli decomposition in Fig.~\ref{fig:twoqubit}(c) shows a nonzero nonlocal block: at the operating point $\|h_{\alpha\beta}\|=0.22~\mu\mathrm{eV}$, with $h_{zz}=-0.04~\mu\mathrm{eV}$ and $(h_{xz}^2+h_{yz}^2)^{1/2}=(h_{zx}^2+h_{zy}^2)^{1/2}=0.12~\mu\mathrm{eV}$. This first-order projection omits virtual non-logical CI states, which raise the effective norm to $\approx1.0~\mu$eV~\cite{Note}. The block stays finite for $\epsilon=0$--$6~\mathrm{meV}$ and decays rapidly with dot separation.

Fig.~\ref{fig:twoqubit}(d) checks that the driven motion is coherent two-qubit evolution: near half transfer, nearly equal $|01\rangle_{\rm loc}$ and $|10\rangle_{\rm loc}$ populations with $C_{\rm log}\approx1$ identify a coherent Bell-like superposition, and the transfer to $|01\rangle_{\rm loc}$ then nearly completes with negligible leakage out of the CI manifold.

\rev{Applying the same pulse to all four logical states gives the full process~\cite{Note}: at half the pulse the dressed logical block, in single-qubit frames, matches an fSim$(\theta,\varphi)$ gate~\cite{Foxen2020} with $\theta\simeq\pi/4$ and a conditional phase $\varphi=5.1$~rad after virtual-$Z$ corrections, a perfect entangler~\cite{Zhang2003}. The gate is slow ($\approx10$~ns) because the static nonlocal block is small; CZ/CNOT calibration, crosstalk, and noise during two-qubit operation are beyond the present scope.}

\textit{Material considerations.}---\rev{Equation~(\ref{eq:continuum_band}) identifies the material inputs, which are standard first-principles outputs: the band mass $m=\hbar^2/2A$ and $d$-wave coefficient $\lambda$ at a spin-degenerate band edge at $\Gamma$, and $g_\perp/g$. Our parameters correspond to $m\approx0.5\,m_e$ and $\lambda/A=0.02$--$0.04$. For the $\Gamma$-point conduction band of the layered $d$-wave altermagnetic semiconductor La$_2$O$_3$Mn$_2$Se$_2$, first-principles $k\cdot p$ fits give $m=0.42\,m_e$ and $\lambda/A=0.54$~\cite{Zhai2025Strain}. A lattice model fitted to these values~\cite{Note} yields a 30 times larger quadrupolar matrix element, so our minimal parameters are conservative for the drive strength. The shape sensitivity away from the circular point is correspondingly larger. At the circular sweet spot, dephasing through $\Delta_{\rm AM}$ becomes second order, while field and exchange noise stay first order through $g_\perp$. There, at $B_x=0.2$~T ($f_q=4.8$~GHz), $X_{\pi/2}$ and $X_\pi$ gates of 1.7 and 3.4~ns need a drive amplitude of only $0.06\%$ of the mean curvature. For $\sigma_\mu=1.3$--$4.3~\mu$eV, charge noise limits $T_2^*$ to $24$--$206~\mu$s and, with its $1/f$ spectrum extrapolated to $f_q$, gives $T_1\approx26$--$290~\mu$s through the transverse quadrupolar coupling, while field noise gives $T_2^*\approx9.5~\mu$s at $\sigma_B=1~\mu$T~\cite{Note}.}

In a proximity design the altermagnet supplies the $d$-wave splitting to a separate clean channel and may itself be metallic; in a standalone design it must also host the confined electron and should be semiconducting or insulating. In both cases a large zero-nuclear-spin isotope abundance reduces hyperfine noise. \rev{The experimentally reported $d$-wave altermagnets KV$_2$Se$_2$O~\cite{Jiang2025KV2Se2O} and Rb$_{1-\delta}$V$_2$Te$_2$O~\cite{Zhang2025RbV2Te2O} are metallic and could serve as proximity layers, with vanadium hyperfine coupling reduced by the small weight of the electron in the altermagnetic layer. The altermagnetism of RuO$_2$ is disputed: film, transport, and spectroscopic studies report signatures~\cite{Feng2022,Bose2022,Karube2022,Fedchenko2024,He2025RuO2}, whereas bulk muon spin rotation finds no magnetic order~\cite{Hiraishi2024RuO2,Tsirlin2026RuO2}.} \rev{Fe$_2$Se$_2$O and related M$_2$X$_2$O monolayers are theoretical candidates only, with functional-dependent electronic structures ranging from a small gap~\cite{Wu2024Fe2Se2O} to a possible Weyl semimetal~\cite{Zou2024M2X2O} and band edges at the spin-valley-locked X and Y points, a regime different from the $\Gamma$-centered one studied here.}

\rev{\textit{Open challenges.}---These are idealized model calculations, and a realization faces several bottlenecks. (i) The SOC-free drive needs a static in-plane field of $0.2$--$0.5$~T, as used for silicon spin qubits; perpendicular to $\mathbf{L}$ it causes no spin flop, but it sets the qubit frequency, adds field noise ($\partial f_q/\partial B_x=20$~GHz/T), must be compatible with superconducting resonators, and renormalizes $g_\perp$ through canting~\cite{Note}; with interfacial Rashba coupling, zero-field EDSR is an alternative. A single-spin route free of both field and SOC needs noncollinear order, e.g., a texture $\lambda[(k_x^2-k_y^2)\sigma_z+2k_xk_y\sigma_x]$, which arises in $120^\circ$ kagome antiferromagnets~\cite{Lee2024KagomeTexture}; related exchange-driven textures have been measured in noncoplanar MnTe$_2$~\cite{Zhu2024MnTe2}. A suitable semiconducting host remains to be identified. (ii) No gate-defined dot has been realized in an altermagnet; this needs a semiconducting altermagnet with a $\Gamma$-point band edge, or a proximitized channel, with low interface disorder (1~meV rms disorder shifts $f_q$ by $\approx6\%$ but preserves the doublet and drive~\cite{Note}), and a dot within a single N\'eel domain, since a domain reversal reverses $\Delta_{\rm AM}$. (iii) Shear strain with the symmetry of the order parameter acts as an effective field along $\mathbf{L}$~\cite{Zhai2025Strain}; static strain must be compensated by the gate-tunable $\Delta_{\rm AM}$, and phonons relax the qubit through the same coupling, which favors low qubit frequencies and hosts with weak spin--strain coupling~\cite{Note}. (iv) Readout, initialization, the magnon bath (gapped by anisotropy and frozen out at millikelvin temperatures), and multi-qubit operation under noise remain to be studied.}

\textit{Conclusion.}---We have proposed a gate-defined spin-qubit platform based on elliptically confined altermagnetic QDs. The central mechanism is that confinement projects the $d$-wave altermagnetic spin splitting into the discrete QD spectrum, producing a low-energy doublet with opposite spin polarization. A phase-controlled resonant quadrupolar gate drive can then modulate this projected altermagnetic field and generate $X_{\pi/2}$ and $X_\pi$ rotations with low leakage in the absence of SOC. This constitutes an electrical spin-drive channel distinct from EDSR, micromagnet gradients, and SOC-derived $g$-tensor modulation.

\rev{An effective-mass theory reproduces the atomistic spectra, the qubit has no first-order coupling to uniform electric-field noise, and its first-order quadrupolar sensitivity vanishes at the circular point, where $T_2^*$ limited by typical charge noise reaches $\sim100$~ms, compared with microseconds at the elliptical gate point (minimal-model parameters).} For two electrons, the same atomistic Hamiltonian supports a localized two-qubit manifold whose projection onto isolated-dot products carries a nonzero nonlocal Pauli block, and resonant barrier modulation implements a perfectly entangling fSim gate. With the phase-controlled rotations and virtual-$Z$ updates, this provides the ingredients for universal gate-based control~\cite{Burkard2023,Bremner2002} within the model---without SOC, micromagnet gradients, or a net magnetic moment. \rev{Its realization hinges on the material and device conditions listed above; the next steps are Wannier-based atomistic models of candidate materials combined with realistic gate electrostatics and time-correlated noise.}

\textit{Note added.}---After the first version of this manuscript was posted, an independent proposal for spin qubits in altermagnetic quantum dots appeared~\cite{AbadilloUriel2026}. That work uses a continuum effective-mass description with zero-field Rashba EDSR for single-qubit control and also discusses singlet-triplet encodings. It is complementary to the present atomistic, sublattice-resolved study, whose quadrupolar drive operates without SOC; the two works agree on the gate-geometric tunability of the confinement-projected altermagnetic splitting.

\textit{Acknowledgments.}---
This work was supported by the National Science Foundation/EPSCoR RII Track-1: Emergent Quantum Materials and Technologies (EQUATE), Award OIA-2044049 and by the UNL Grand Challenges catalyst award entitled Quantum Approaches addressing Global Threats. The calculations and writing of the paper were reviewed by Claude Opus 5.5.

\pdfbookmark[1]{References}{references}
\let\SavedAddcontentsline\addcontentsline
\renewcommand{\addcontentsline}[3]{}
\bibliography{main}
\let\addcontentsline\SavedAddcontentsline

\clearpage
\onecolumngrid
\setcounter{secnumdepth}{2}
\setcounter{section}{0}
\setcounter{equation}{0}
\setcounter{figure}{0}
\setcounter{table}{0}
\renewcommand{\thesection}{S\arabic{section}}
\renewcommand{\thesubsection}{\thesection.\arabic{subsection}}
\renewcommand{\theequation}{S\arabic{equation}}
\renewcommand{\thefigure}{S\arabic{figure}}
\renewcommand{\thetable}{S\arabic{table}}
\renewcommand{\theHsection}{S\arabic{section}}
\renewcommand{\theHequation}{S\arabic{equation}}
\renewcommand{\theHfigure}{S\arabic{figure}}
\renewcommand{\theHtable}{S\arabic{table}}

\pdfbookmark[1]{Supplemental Material}{supplemental}
\begin{center}
{\large\bfseries Supplemental Material: Gate-Controlled Spin Qubits in Confined Altermagnets}\\[0.6em]
Hamed Vakili
\end{center}

\tableofcontents

\section{Hamiltonian, parameters, and numerical setup}

The atomistic model is the second-quantized checkerboard Hamiltonian of the main text. Each site \(i\) is one atom, \(c_{is}^{\dagger}\) creates an electron with spin \(s=\uparrow,\downarrow\) on that atom, \(n_{is}=c_{is}^\dagger c_{is}\), and the sublattice label is \(\eta_i=+1\) on \(A\) and \(-1\) on \(B\). The supplementary scans use
\begin{align}
H_{\rm 1e}^{\rm supp}={}&
H_{\rm hop}+H_{\rm ex}+H_Z+H_V+H_R,
\nonumber\\
H_{\rm hop}={}&
\Delta_{ab}\sum_{\langle ij\rangle,s}
\left(c_{is}^{\dagger}c_{js}+{\rm H.c.}\right)
+t_1\sum_{\langle\langle ij\rangle\rangle,s}
\left(c_{is}^{\dagger}c_{js}+{\rm H.c.}\right)
\nonumber\\
&+\sum_{i,s}
\left[
(t_2+\eta_i t_{AM})c_{is}^{\dagger}c_{i+2\hat{x},s}
+(t_2-\eta_i t_{AM})c_{is}^{\dagger}c_{i+2\hat{y},s}
+{\rm H.c.}
\right]
-\mu\sum_{i,s}n_{is},
\nonumber\\
H_R={}&
t_R\sum_i
\left[
c_{i\uparrow}^{\dagger}c_{i+\hat{x},\downarrow}
-c_{i\downarrow}^{\dagger}c_{i+\hat{x},\uparrow}
-i c_{i\uparrow}^{\dagger}c_{i+\hat{y},\downarrow}
-i c_{i\downarrow}^{\dagger}c_{i+\hat{y},\uparrow}
+{\rm H.c.}
\right],
\qquad
t_R=\frac{\alpha_R}{2a}.
\label{eq:supp_hamiltonian}
\end{align}
Here \(\langle ij\rangle\) denotes cardinal nearest-neighbor \(A\)-\(B\) bonds and \(\langle\langle ij\rangle\rangle\) diagonal same-sublattice bonds. The exchange \(H_{\rm ex}=\tfrac{J}{2}\sum_i\eta_ic_i^\dagger(\mathbf L\cdot\bm\sigma)c_i\) with \(\mathbf L\parallel\hat z\), the Zeeman term \(H_Z=\tfrac12g\mu_BB_x\sum_ic_i^\dagger\sigma_xc_i\) with \(g=2\), and the confinement \(H_V=\sum_{i,s}(V_xx_i^2+V_yy_i^2)n_{is}\) are those of the main text. The Rashba term \(H_R\), a central-difference discretization of \(\alpha_R(k_y\sigma_x-k_x\sigma_y)\), is included only in Secs.~\ref{sec:cuts}--\ref{sec:edsr}.

All single-dot calculations use \(a=0.5\)~nm and a converged \(L=40\)~nm lattice (Sec.~\ref{sec:convergence}). The two main-text parameter sets are
\begin{itemize}
\item Fig.~2 set: \(t_1=-150\)~meV, \(t_2=0\), \(t_{AM}=20\)~meV, \(\Delta_{ab}=8\)~meV, \(J=10\)~meV, \(\mu=0\), \(\bar V=0.13~{\rm meV\,nm^{-2}}\); representative point \(\delta=0.3\) [\((V_x,V_y)=(0.091,0.169)~{\rm meV\,nm^{-2}}\)], \(B_x=0.3\)~T.
\item Fig.~3 set: \(t_1=-150\)~meV, \(t_2=0\), \(t_{AM}=15\)~meV, \(\Delta_{ab}=12\)~meV, \(J=12\)~meV, \(\mu=0\), \((V_x,V_y)=(0.18,0.43)~{\rm meV\,nm^{-2}}\) (\(\delta=0.41\)), \(B_x=0.45\)~T.
\end{itemize}
The scans of Secs.~\ref{sec:cuts} and~\ref{sec:maps} vary one or two parameters around the Fig.~2 set at its representative point. The quadrupolar drive is \(H_d(t)=A_Q\cos(\omega_qt+\phi)Q\) with \(Q=x^2-y^2\); its rotating-wave Rabi frequency is \(f_R=A_Q|\langle1|Q|0\rangle|/h\), quoted for the reference amplitude \(A_Q=10^{-5}~{\rm eV\,nm^{-2}}\) unless stated otherwise and linear in \(A_Q\). The qubit and leakage gaps are \(\Delta_q=E_1-E_0\) and \(\Delta_{\rm leak}=E_2-E_1\) (\(E_2-E_0\) where stated).

\section{Continuum limit and closed-form dot spectrum}
\label{sec:continuum}

\subsection{Lowest band near \texorpdfstring{$\Gamma$}{Gamma}}
In the two-site unit cell and for spin \(\sigma=\pm1\) along \(\mathbf L\), the Bloch Hamiltonian of Eq.~(\ref{eq:supp_hamiltonian}) without \(H_R\) is
\begin{equation}
H_\sigma(\mathbf k)=\varepsilon_0(\mathbf k)+\big[d(\mathbf k)+\sigma J/2\big]\tau_z+\gamma(\mathbf k)\tau_x,
\label{eq:bloch}
\end{equation}
with sublattice Pauli matrices \(\tau\), \(\varepsilon_0=4t_1\cos k_xa\cos k_ya+2t_2(\cos2k_xa+\cos2k_ya)\), \(d=2t_{AM}(\cos2k_xa-\cos2k_ya)\), and \(\gamma=2\Delta_{ab}(\cos k_xa+\cos k_ya)\). The lower band is \(E_{-,\sigma}=\varepsilon_0-[(d+\sigma J/2)^2+\gamma^2]^{1/2}\). Expanding to second order in \(\mathbf k\) around \(\Gamma\) (where \(d=0\) and \(\gamma=4\Delta_{ab}\)) for \(t_2=0\) gives Eq.~(3) of the main text,
\begin{equation}
E_\sigma(\mathbf k)\simeq E_{\min}+(A+\sigma\lambda)k_x^2+(A-\sigma\lambda)k_y^2,\quad
A=a^2\Big(\frac{4\Delta_{ab}^2}{R_0}-2t_1\Big),\quad
\lambda=\frac{2Jt_{AM}a^2}{R_0},
\label{eq:supp_band}
\end{equation}
with \(R_0=[(J/2)^2+16\Delta_{ab}^2]^{1/2}\) and \(E_{\min}=4t_1-R_0\). The \(d\)-wave term acts on the lower band only through its exchange-induced sublattice polarization, \(\langle\tau_z\rangle=-\sigma J/2R_0\). The two bands at \(\Gamma\) are separated by \(2R_0\), which is \(65\)~meV (Fig.~2 set) and \(97\)~meV (Fig.~3 set), far above all dot energies.

\subsection{Dot levels and field-free splitting}
For \(V=V_xx^2+V_yy^2\), each direction of Eq.~(\ref{eq:supp_band}) is a harmonic oscillator, \(\hbar\omega=2[(A\pm\lambda)V_{x,y}]^{1/2}\), with ground-state width \(\ell^4=(A\pm\lambda)/V_{x,y}\). Hence
\begin{equation}
E_{n\sigma}=E_{\min}+2\sqrt{(A+\sigma\lambda)V_x}\Big(n_x+\frac12\Big)+2\sqrt{(A-\sigma\lambda)V_y}\Big(n_y+\frac12\Big),
\end{equation}
and the splitting of the ground doublet is
\begin{equation}
\Delta_{\rm AM}\equiv E_{0\downarrow}-E_{0\uparrow}=\big(\sqrt{A+\lambda}-\sqrt{A-\lambda}\big)\big(\sqrt{V_y}-\sqrt{V_x}\big).
\label{eq:supp_dam}
\end{equation}
With \(V_{x,y}=\bar V(1\mp\delta)\) and \(\sqrt{1+\delta}-\sqrt{1-\delta}=\delta+\delta^3/8+O(\delta^5)\), Eq.~(\ref{eq:supp_dam}) reduces to Eq.~(4) of the main text. The splitting is therefore linear in \(\delta\), proportional to \(\sqrt{\bar V}\), and proportional to \(\lambda\propto Jt_{AM}/R_0\) for \(\lambda\ll A\). This explains the trends of the parameter scans in Secs.~\ref{sec:cuts} and~\ref{sec:maps}: \(\Delta_q\) grows with \(J\) and \(t_{AM}\) and decreases with \(\Delta_{ab}\).

\subsection{Two-level description, transverse factor, and quadrupole matrix elements}
The in-plane field couples \(|\!\uparrow,\phi_\uparrow\rangle\) and \(|\!\downarrow,\phi_\downarrow\rangle\) with strength \(\tfrac12g\mu_BB_x\langle u_\uparrow|u_\downarrow\rangle\langle\phi_\uparrow|\phi_\downarrow\rangle\), where \(u_\sigma\) is the sublattice spinor of the lower band at \(\Gamma\) and \(\phi_\sigma\) the anisotropic oscillator ground state. The doublet is thus described by \(H_q=-\tfrac12\Delta_{\rm AM}\sigma_z+\tfrac12g_\perp\mu_BB_x\sigma_x\) with \(g_\perp=g\,\eta_{\rm sub}\eta_{\rm orb}\),
\begin{equation}
\eta_{\rm sub}=\frac{4\Delta_{ab}}{R_0},\qquad
\eta_{\rm orb}=\prod_{i=x,y}\Big[\frac{2\ell_{i\uparrow}\ell_{i\downarrow}}{\ell_{i\uparrow}^2+\ell_{i\downarrow}^2}\Big]^{1/2},
\end{equation}
\(\ell_{x\uparrow}^4=(A+\lambda)/V_x\), \(\ell_{y\uparrow}^4=(A-\lambda)/V_y\), and \(\lambda\to-\lambda\) for spin down. For both main-text sets \(\eta_{\rm sub}=0.988\) and \(0.992\), and \(\eta_{\rm orb}>0.9997\). Accordingly, \(E_1-E_0=[\Delta_{\rm AM}^2+(g_\perp\mu_BB_x)^2]^{1/2}\) and \(\langle\sigma_z\rangle_{0,1}=\pm\cos\theta\) with \(\tan\theta=g_\perp\mu_BB_x/\Delta_{\rm AM}\). For \(\Delta_{ab}\to0\), \(\eta_{\rm sub}\to0\): the two spin species reside on opposite sublattices, the field cannot mix them, and the quadrupolar drive vanishes (Sec.~\ref{sec:cuts}).

Because \(Q\) is spin independent, its matrix elements follow from the spin-resolved quadrupole moments \(Q_\sigma=\langle x^2\rangle_\sigma-\langle y^2\rangle_\sigma\), with \(\langle x^2\rangle=\ell_x^2/2\),
\begin{align}
Q_\uparrow-Q_\downarrow&=\tfrac12\big(\sqrt{A+\lambda}-\sqrt{A-\lambda}\big)\big(V_x^{-1/2}+V_y^{-1/2}\big),
\nonumber\\
|\langle1|Q|0\rangle|&=\tfrac12|Q_\uparrow-Q_\downarrow|\sin\theta,\qquad
|Q_{11}-Q_{00}|=|Q_\uparrow-Q_\downarrow|\cos\theta.
\label{eq:supp_q}
\end{align}
The transverse (driving) and longitudinal (dephasing) parts of the drive operator are therefore in the fixed ratio \(|Q_{11}-Q_{00}|/2|Q_{01}|=\cot\theta\). Equivalently, \(Q_\uparrow-Q_\downarrow=-\partial\Delta_{\rm AM}/\partial A_Q\) for \(V_{x,y}\to V_{x,y}\pm A_Q\).

\subsection{Comparison with the atomistic model}
Table~\ref{tab:continuum} compares the closed-form expressions with the atomistic lattice (\(L=40\)~nm) at the two operating points. Over the full ellipticity range of main-text Fig.~2(a), the atomistic \(\Delta_{\rm AM}\) and \(E_1-E_0\) agree with the continuum expressions to within \(0.6\%\) for \(|\delta|\ge0.1\); the small remaining difference comes from band nonparabolicity.

\begin{table}[!htbp]
\caption{Closed-form continuum expressions (Sec.~\ref{sec:continuum}) versus the atomistic lattice model at the main-text operating points (\(L=40\)~nm).}
\label{tab:continuum}
\begin{ruledtabular}
\begin{tabular}{lcccc}
& \multicolumn{2}{c}{Fig.~2 point (\(\delta=0.3\), 0.3~T)} & \multicolumn{2}{c}{Fig.~3 point (\(\delta=0.41\), 0.45~T)}\\
quantity & atomistic & continuum & atomistic & continuum\\
\hline
\(A\) (meV\,nm\(^2\)) / band mass & \multicolumn{2}{c}{77.0 / \(0.495\,m_e\)} & \multicolumn{2}{c}{78.0 / \(0.489\,m_e\)}\\
\(\lambda\) (meV\,nm\(^2\)) / \(2R_0\) (meV) & \multicolumn{2}{c}{3.09 / 64.8} & \multicolumn{2}{c}{1.86 / 96.7}\\
\(\Delta_{\rm AM}\) (\(\mu\)eV) & 38.72 & 38.52 & 48.86 & 48.77\\
\(E_1-E_0\) (\(\mu\)eV) & 51.73 & 51.58 & 71.12 & 71.07\\
\(|\langle\sigma_z\rangle_{0,1}|\) & 0.749 & 0.747 & 0.687 & 0.686\\
\(|\langle1|Q|0\rangle|\) (nm\(^2\)) & 0.3370 & 0.3364 & 0.1489 & 0.1487\\
\(|Q_{11}-Q_{00}|\) (nm\(^2\)) & 0.7607 & 0.7554 & 0.2814 & 0.2807\\
\end{tabular}
\end{ruledtabular}
\end{table}

\section{Numerical convergence}
\label{sec:convergence}

\subsection{Box size}
\begin{figure}[!htbp]
\centering
\includegraphics[width=0.96\textwidth]{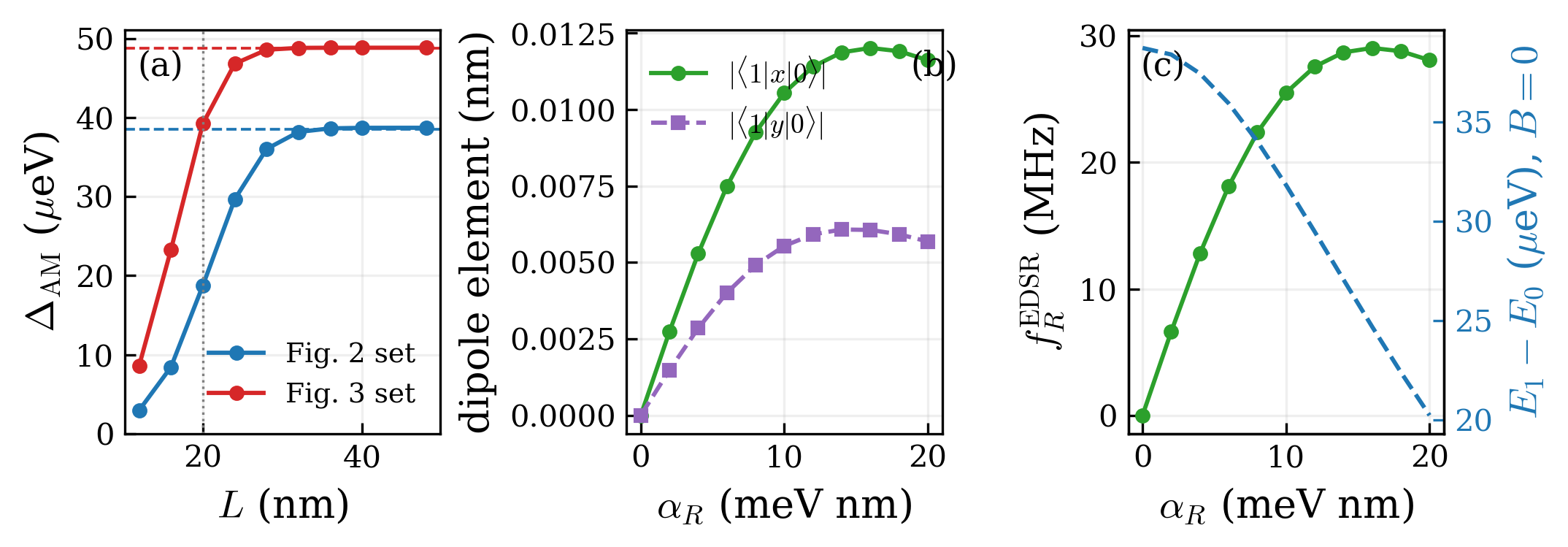}
\caption{(a) Field-free splitting \(\Delta_{\rm AM}\) versus box size \(L\) for the Fig.~2 set (\(\delta=0.3\)) and the Fig.~3 set; dashed lines are Eq.~(\ref{eq:supp_dam}), and the dotted line marks \(L=20\)~nm. (b) Zero-applied-field EDSR dipole matrix elements \(|\langle1|x|0\rangle|\) and \(|\langle1|y|0\rangle|\) versus interfacial Rashba coupling \(\alpha_R\) for the Fig.~2 set at \(\delta=0.3\), \(B=0\), \(L=40\)~nm. (c) Rotating-wave EDSR Rabi frequency \(f_R^{\rm EDSR}=A_x|\langle1|x|0\rangle|/h\) for \(A_x=10^{-5}~{\rm eV\,nm^{-1}}\) and the zero-field qubit gap.}
\label{fig:box_edsr}
\end{figure}

The open-boundary lattice adds hard-wall confinement. Because \(\Delta_{\rm AM}\propto\langle k_y^2\rangle-\langle k_x^2\rangle\) is a small difference of two kinetic energies, a box that clips the weaker confinement direction raises \(\langle k_x^2\rangle\), makes the dot effectively more circular, and strongly reduces the splitting. Figure~\ref{fig:box_edsr}(a) shows that \(\Delta_{\rm AM}\) converges to the continuum value of Eq.~(\ref{eq:supp_dam}) for \(L\gtrsim36\)~nm. At \(L=20\)~nm, \(\Delta_{\rm AM}\) is \(18.8~\mu\)eV instead of \(38.7~\mu\)eV (Fig.~2 set) and \(39.2~\mu\)eV instead of \(48.9~\mu\)eV (Fig.~3 set), and the leakage gap \(E_2-E_0\) is overestimated (6.4 versus 5.2~meV for the Fig.~2 set). At \(L=40\)~nm the remaining change on going to \(L=48\)~nm is \(<0.02\%\) at the representative points and \(<0.3\%\) at \(|\delta|=0.65\). The two-electron calculations use \(L=60\)~nm with dot centers \(\approx21\)~nm from the boundaries along \(x\), equivalent to the converged single-dot box.

\subsection{Time step, basis size, and pulse length of the single-qubit gates}
The driven evolution uses stepwise midpoint matrix exponentials in the basis of the lowest 64 eigenstates at the operating point, with an integer number of steps per carrier period. Table~\ref{tab:numerics} shows the \(X_\pi\) average gate fidelity, maximized over the equatorial rotation axis, and the maximal leakage for several time steps and pulse lengths. A time step of 3~ps produces spurious leakage of order \(10^{-3}\)--\(10^{-2}\) through the poorly resolved high-energy projected states; for \(\le1\)~ps the leakage is \(\lesssim10^{-5}\) and the remaining infidelity is a coherent error from the counter-rotating (Bloch--Siegert) term and off-resonant orbital levels that decreases with pulse length. The main-text gates use 8 and 16 carrier cycles and a 0.5~ps step; halving the step changes \(\bar F(X_\pi)\) by \(5\times10^{-7}\) and its leakage from \(0.9\times10^{-6}\) to \(1.2\times10^{-6}\), and enlarging the basis to 96 states changes \(\bar F(X_\pi)\) by \(10^{-10}\). The average gate fidelity is evaluated as \(\bar F=[{\rm Tr}(MM^\dagger)+|{\rm Tr}M|^2]/[d(d+1)]\) with \(M=U_T^\dagger U_L\), where \(U_L\) is the \(d\times d\) logical block of the propagator and \(U_T\) the target, which accounts for leakage~\cite{Pedersen2007}. At an integer number of carrier cycles the laboratory and rotating frames coincide up to a global phase.

\textit{Switch-on phase.}---The pulses have rectangular envelopes, so fidelity and leakage depend on the phase of the drive when it is switched on. Unless stated otherwise, all gates in the main text and here start and end at a node of the drive (\(\phi=\pi/2\) in main-text Eq.~(5)), which minimizes the leakage. At the Fig.~3 point and fixed amplitude, scanning \(\phi\) over \(2\pi\) gives \(\bar F(X_{\pi/2})=0.99927\)--\(0.99967\), \(\bar F(X_\pi)=0.99931\)--\(0.99962\), and \(X_\pi\) leakage from \(9\times10^{-7}\) at a node to \(6.5\times10^{-4}\) near an antinode. Because \(\phi\) also sets the rotating-frame axis, a pulse about another axis keeps a node start by shifting its start time by a fraction of a carrier period (a quarter period from \(X\) to \(Y\)). All 8- and 16-cycle gates here have \(f_R/f_q=1/32\), so at \(\delta=0\) they share the same counter-rotating error, and their fidelities differ only through \(\phi\). A node start gives \(\bar F=0.99982\) (\(X_{\pi/2}\)) and \(0.99963\) (\(X_\pi\)) for both the circular Fig.~3-set dot (Sec.~\ref{sec:noise}) and the DFT sweet spot (Sec.~\ref{sec:dft}). An optimized start phase raises \(\bar F(X_\pi)\) to 0.99992 with leakage \(5\times10^{-5}\) in the former (\(\phi\approx0.8\)~rad) and above 0.9999 with leakage \(<10^{-7}\) in the latter (\(\phi\approx0.5\)~rad), where the RF amplitude is far smaller.

\begin{table}[!htbp]
\caption{Convergence of the \(X_\pi\) gate at the Fig.~3 point, for pulses starting at a drive node: best-axis average gate fidelity / maximal leakage from \(|0\rangle,|1\rangle\).}
\label{tab:numerics}
\begin{ruledtabular}
\begin{tabular}{lccc}
\(X_\pi\) length & \(\Delta t=3\)~ps & \(\Delta t=1\)~ps & \(\Delta t=0.5\)~ps\\
\hline
8 cycles (0.47 ns) & 0.99527 / \(2.4\times10^{-3}\) & 0.99724 / \(3.7\times10^{-7}\) & 0.99725 / \(3.0\times10^{-6}\)\\
16 cycles (0.93 ns) & 0.99840 / \(9.5\times10^{-4}\) & 0.99931 / \(8.7\times10^{-8}\) & 0.99931 / \(9.3\times10^{-7}\)\\
24 cycles (1.40 ns) & 0.98247 / \(1.7\times10^{-2}\) & 0.99969 / \(3.9\times10^{-8}\) & 0.99969 / \(3.1\times10^{-7}\)\\
\end{tabular}
\end{ruledtabular}
\end{table}

\section{One-parameter cuts}
\label{sec:cuts}

\begin{figure}[!htbp]
\centering
\includegraphics[width=0.9\textwidth]{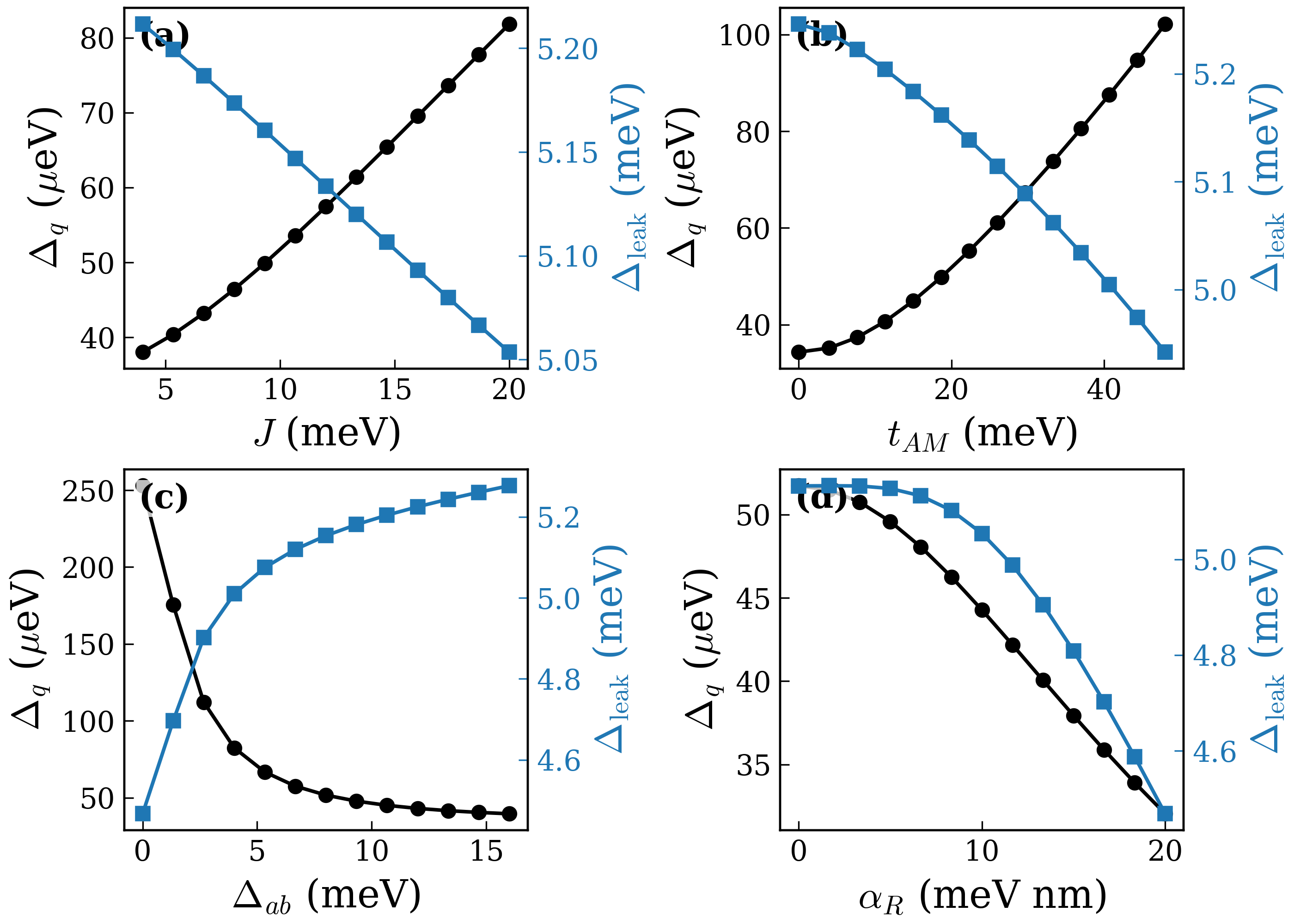}
\caption{Single-parameter scans of the qubit gap \(\Delta_q=E_1-E_0\) (black circles) and leakage gap \(\Delta_{\rm leak}=E_2-E_1\) (blue squares) in the atomistic model (\(L=40\)~nm). In each panel one parameter is varied while the others are fixed at \(t_1=-150\)~meV, \(t_2=0\), \(\mu=0\), \(J=10\)~meV, \(t_{AM}=20\)~meV, \(\Delta_{ab}=8\)~meV, and \(\alpha_R=0\); \(B_x=0.3\)~T and \((V_x,V_y)=(0.091,0.169)~{\rm meV\,nm^{-2}}\).}
\label{fig:supp_gap_leakage}
\end{figure}

\begin{figure}[!htbp]
\centering
\includegraphics[width=0.9\textwidth]{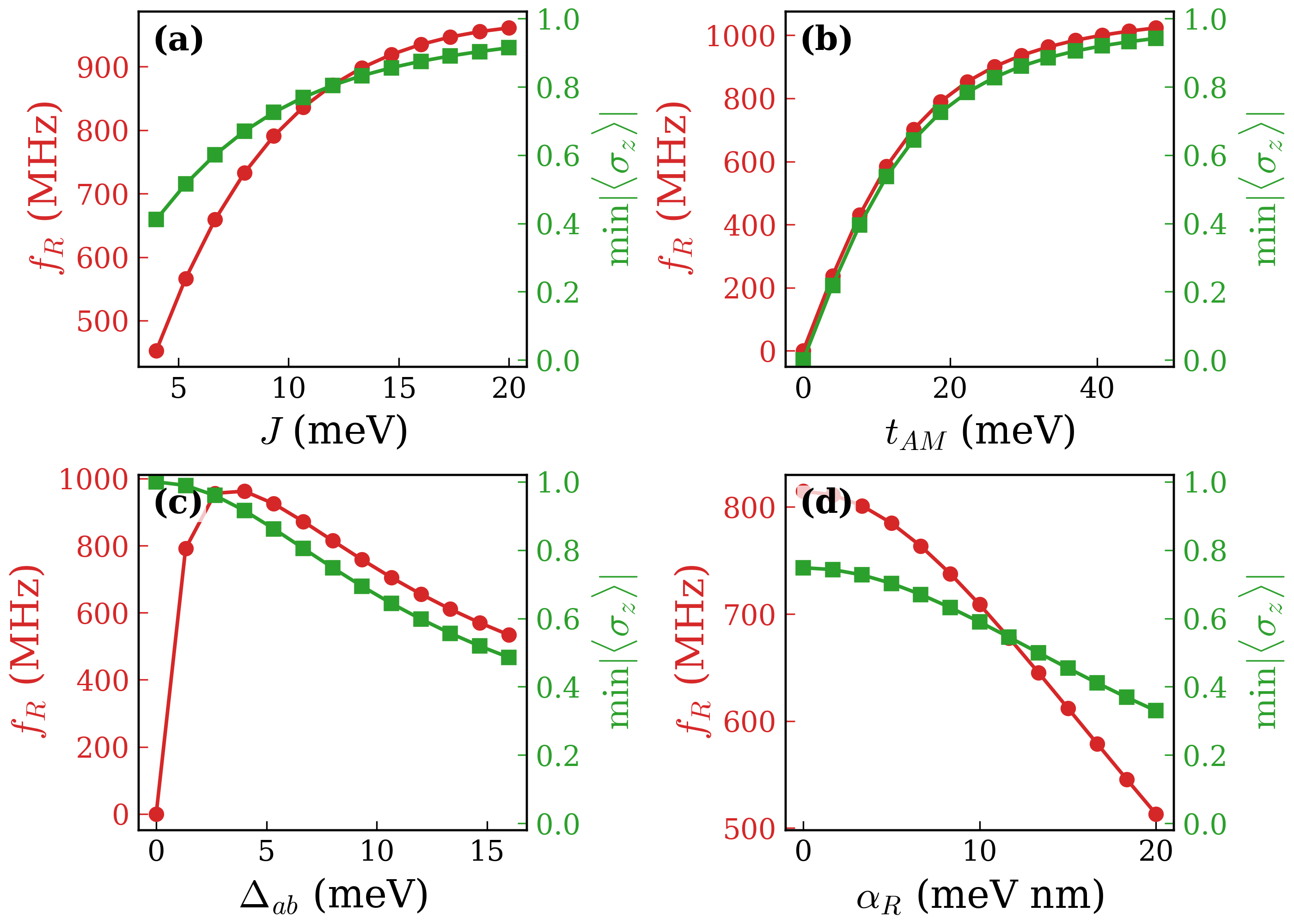}
\caption{Rotating-wave Rabi-rate estimate \(f_R=A_Q|\langle 1|x^2-y^2|0\rangle|/h\) for \(A_Q=10^{-5}~{\rm eV\,nm^{-2}}\) (red circles) and polarization \(\min(|\langle\sigma_z\rangle_0|,|\langle\sigma_z\rangle_1|)\) (green squares) for the scans of Fig.~\ref{fig:supp_gap_leakage}.}
\label{fig:supp_rate_pol}
\end{figure}

Figures~\ref{fig:supp_gap_leakage} and~\ref{fig:supp_rate_pol} show the qubit and leakage gaps, the Rabi-rate estimate, and the logical-state polarization \({\rm Pol.}=\min(|\langle\sigma_z\rangle_0|,|\langle\sigma_z\rangle_1|)\) when one parameter is varied at a time around the Fig.~2 set (\(\delta=0.3\), \(B_x=0.3\)~T, \(L=40\)~nm). The qubit gap remains in the microwave range and the leakage gap in the meV range throughout. The trends follow Sec.~\ref{sec:continuum}: increasing \(J\) or \(t_{AM}\) increases \(\lambda\) and thus \(\Delta_{\rm AM}\), which raises both the polarization \(\cos\theta\) and, through \(|Q_\uparrow-Q_\downarrow|\), the drive matrix element. At \(\Delta_{ab}=0\) the spectrum remains split but the quadrupolar matrix element vanishes exactly. With \(\Delta_{ab}=0\) the \(A\) and \(B\) sublattices decouple orbitally (the \(t_1\), \(t_2\), and \(t_{AM}\) bonds are all same-sublattice), so the two spin species of the doublet live on opposite sublattices (\(\eta_{\rm sub}=0\)), the field cannot mix them, and the spin-independent, site-diagonal \(Q\) cannot connect them. Increasing \(\Delta_{ab}\) beyond a few meV reduces \(\lambda\propto1/R_0\) and therefore the splitting and the rate. Interfacial Rashba coupling reduces the gap, rate, and polarization at large \(\alpha_R\) (at this field the polarization falls below 0.6 for \(\alpha_R\gtrsim10\)~meV\,nm), but the gaps and the drive persist over the plotted range, which supports using the SOC-free limit as the clean reference point.

\section{Two-parameter maps}
\label{sec:maps}

\begin{figure}[!htbp]
\centering
\includegraphics[width=0.9\textwidth]{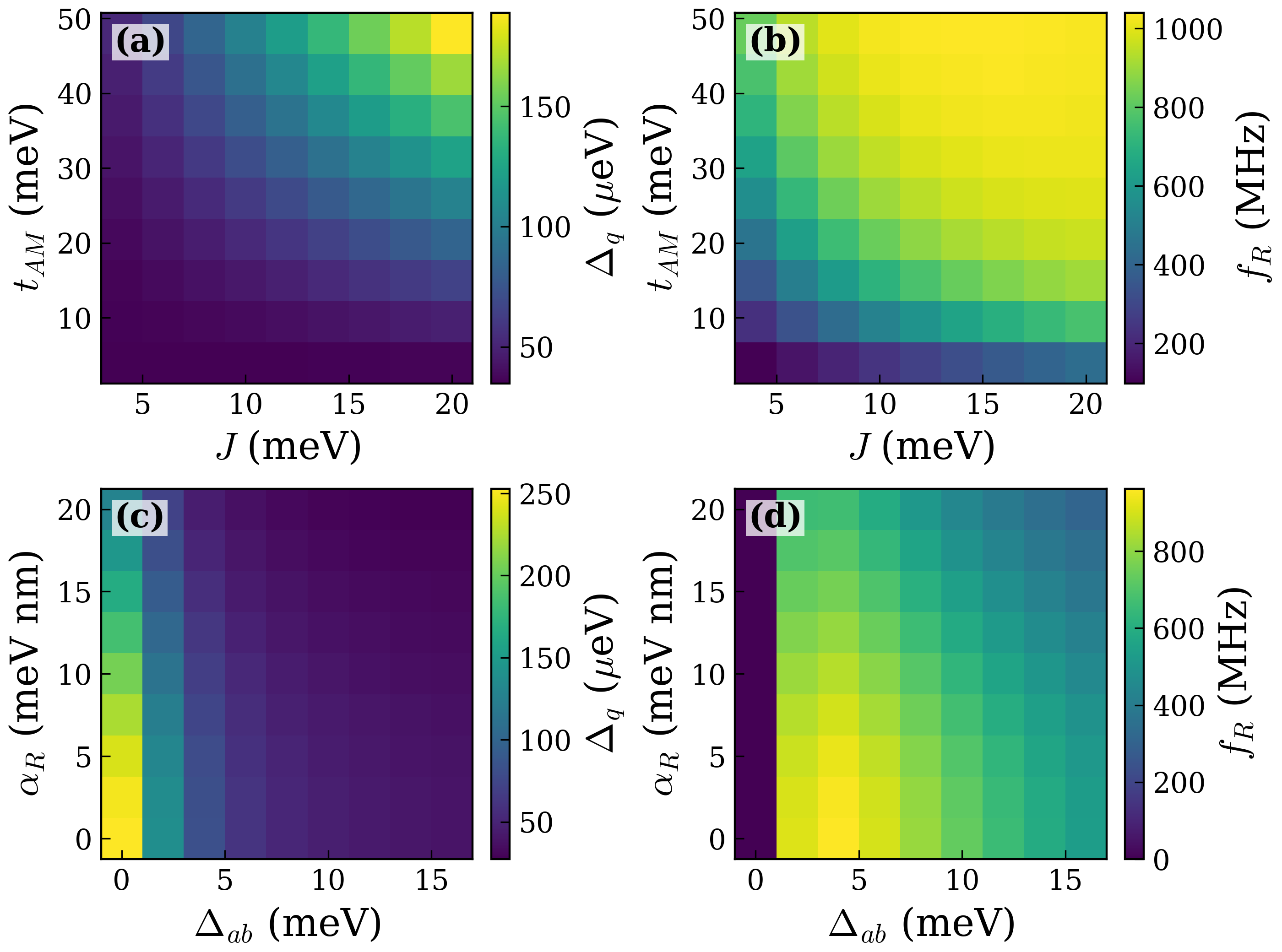}
\caption{Two-parameter scans (\(L=40\)~nm, \(B_x=0.3\)~T, \((V_x,V_y)=(0.091,0.169)~{\rm meV\,nm^{-2}}\)). (a,b) Qubit gap and Rabi-rate estimate versus \(J\) and \(t_{AM}\), with \(\Delta_{ab}=8\)~meV and \(\alpha_R=0\). (c,d) Qubit gap and Rabi-rate estimate versus \(\Delta_{ab}\) and \(\alpha_R\), with \(J=10\)~meV and \(t_{AM}=20\)~meV.}
\label{fig:supp_maps}
\end{figure}

Figure~\ref{fig:supp_maps} summarizes two-parameter scans. Panels (a,b) vary \(J\) and \(t_{AM}\): the gap grows with \(\lambda\propto Jt_{AM}\), and the transition rate saturates once \(\Delta_{\rm AM}\gtrsim g_\perp\mu_BB_x\) because the rate is the product of \(|Q_\uparrow-Q_\downarrow|\propto\lambda\) and \(\sin\theta\). Panels (c,d) vary \(\Delta_{ab}\) and \(\alpha_R\). The line \(\Delta_{ab}=0\) has vanishing quadrupolar rate across the full Rashba range, extending the \(\alpha_R=0\) result of Sec.~\ref{sec:cuts}. For finite \(\Delta_{ab}\), moderate Rashba coupling coexists with a finite \(f_R\). These maps leave a wide window with \(\Delta_q\) of tens of \(\mu\)eV, \(\Delta_{\rm leak}\) in the meV range, and \(f_R\) of hundreds of MHz for the reference amplitude.

\section{Zero-applied-field SOC-assisted EDSR}
\label{sec:edsr}

If interfacial Rashba coupling is present, the same dot also permits a conventional electric-dipole drive at zero applied field. We diagnose this channel by the dipole matrix elements \(\xi_{x,y}=|\langle1|x,y|0\rangle|\) of the linear drive \(H_{\rm EDSR}(t)=A_xx\cos(\omega_qt)\), whose rotating-wave Rabi frequency is \(f_R^{\rm EDSR}=A_x\xi_x/h\). Figure~\ref{fig:box_edsr}(b,c) shows the result for the Fig.~2 set at \(B=0\). At \(\alpha_R=0\) the dipole element vanishes (numerically \(<10^{-12}\)~nm), as required by in-plane parity in a centered dot. It grows with \(\alpha_R\) to \(\xi_x=0.011\)~nm at \(\alpha_R=10\)--\(20\)~meV\,nm, corresponding to \(f_R^{\rm EDSR}\approx26\)--\(28\)~MHz for \(A_x=10^{-5}~{\rm eV\,nm^{-1}}\), while the zero-field gap decreases from \(38.7\) to \(20.2~\mu\)eV. At \(B=0\) the quadrupolar matrix element remains zero for all \(\alpha_R\) (\(<10^{-9}~{\rm nm^2}\)), so the dipolar (SOC) and quadrupolar (altermagnetic) channels are distinguished by the parity of the drive and by the need for a transverse field.

\section{Static in-plane field: drive--polarization trade-off and N\'eel-vector canting}
\label{sec:field}

\textit{Trade-off.}---Within the two-level description of Sec.~\ref{sec:continuum}, \(B_x\) sets the mixing angle \(\theta\). The quadrupolar Rabi rate is \(\propto\sin\theta\), the logical-state polarization is \(\cos\theta\), and the first-order sensitivity to shape (curvature) noise is \(\propto\cos\theta\) [Eq.~(\ref{eq:supp_q})]. Increasing \(B_x\) therefore increases the drive efficiency and suppresses quadrupolar dephasing at the expense of the spin-sublattice polarization of the logical states, with a ratio of drive to dephasing coupling of \(\tan\theta=g_\perp\mu_BB_x/\Delta_{\rm AM}\). Main-text Fig.~2(b) shows this for the Fig.~2 set, and Fig.~\ref{fig:noise}(c) for the Fig.~3 point. From \(B_x=0.05\) to \(1\)~T, \(f_R\) (reference amplitude) increases from 58 to 456~MHz, \(T_2^*\) for \(\sigma_\mu=1~\mu\)eV increases from 1.8 to \(4.5~\mu\)s, and \(|\langle\sigma_z\rangle|\) decreases from 0.99 to 0.39. The operating contour of main-text Fig.~2(d), \(\cos\theta=0.6\), corresponds to \(g_\perp\mu_BB_x=\tfrac43|\Delta_{\rm AM}|\).

\textit{Canting.}---The field is applied perpendicular to \(\mathbf L\), so no spin flop occurs. It cants both sublattice moments toward \(\hat x\) by an angle \(\varphi\simeq B_x/2B_E\), where \(B_E\) is the exchange field. With \(g\mu_BB_E\sim k_BT_N\) up to factors of order unity, \(\varphi\sim g\mu_BB_x/2k_BT_N\approx10^{-3}\) for \(B_x=0.45\)~T and \(T_N\approx300\)~K. In the exchange term, \(\eta_i\mathbf L\to\eta_i\cos\varphi\,\hat z+\sin\varphi\,\hat x\), which adds a uniform term \(\tfrac12J\sin\varphi\sum_ic_i^\dagger\sigma_xc_i\) parallel to the Zeeman term. The canting therefore renormalizes the transverse field, \(g_\perp\to g_\perp[1\pm J/(2k_BT_N)]\) with a sign set by the sign of the exchange coupling (about \(\pm20\%\) for \(J=12\)~meV and \(T_N=300\)~K), without changing the mechanism. The longitudinal reduction of the staggered exchange is \(O(\varphi^2)\sim10^{-6}\). Because this renormalization enters only the calibration of \(g_\perp\), we keep \(\mathbf L\) rigid in the simulations. For the DFT-parameterized lattice of Sec.~\ref{sec:dft} (\(J=96.9\)~meV), the same estimate gives \(J/(2k_BT_N)\approx1.9\) at 300~K, so the transverse field could differ from \(g_\perp\mu_BB_x\) by a factor of up to about 3. The results of that section depend on the transverse energy, and the quoted \(B_x\) is its bare-field equivalent.

\textit{Practical aspects.}---Fields of 0.2--0.5~T are comparable to those used in silicon spin-qubit devices~\cite{Burkard2023}. The field must be compatible with superconducting resonators and readout circuitry, and it adds a field-noise channel with \(\partial f_q/\partial B_x=g_\perp\mu_B\sin\theta/h\approx20\)~GHz/T at the Fig.~3 point (Table~\ref{tab:sens}). Where interfacial Rashba coupling is present, zero-field EDSR (Sec.~\ref{sec:edsr}) avoids the field altogether.

\section{Noise sensitivities and dephasing}
\label{sec:noise}

\begin{figure}[!htbp]
\centering
\includegraphics[width=0.9\textwidth]{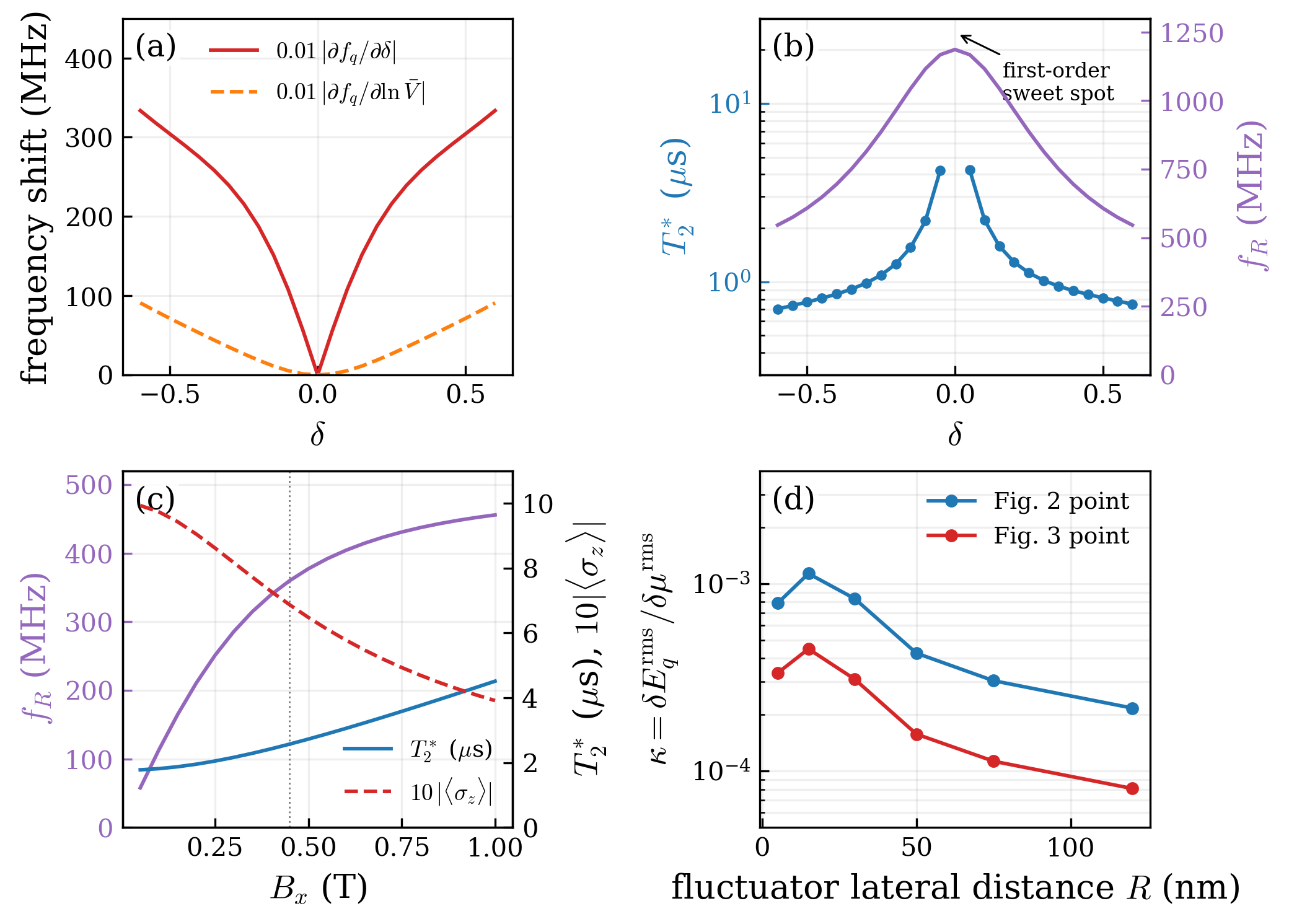}
\caption{(a) First-order shape sensitivities \(0.01\,|\partial f_q/\partial\delta|\) and \(0.01\,|\partial f_q/\partial\ln\bar V|\) versus \(\delta\) (Fig.~2 set, \(B_x=0.3\)~T). (b) \(T_2^*\) for \(\sigma_\mu=1~\mu\)eV from the fluctuator ensemble (left axis) and \(f_R\) for \(A_Q=10^{-5}~{\rm eV\,nm^{-2}}\) (right axis); at \(\delta=0\) the first-order shape sensitivity vanishes and the ensemble value (64~ms) is off scale. (c) \(B_x\) dependence at the Fig.~3 point: \(f_R\), \(T_2^*\), and \(10|\langle\sigma_z\rangle|\); the dotted line marks \(0.45\)~T. (d) Conversion factor \(\kappa\) resolved by the lateral distance of the fluctuators.}
\label{fig:noise}
\end{figure}

\textit{Sensitivities.}---First-order (Hellmann--Feynman) sensitivities of \(E_q=E_1-E_0\) to a parameter \(\lambda_p\) are \(\partial E_q/\partial\lambda_p=\langle1|\partial H/\partial\lambda_p|1\rangle-\langle0|\partial H/\partial\lambda_p|0\rangle\), evaluated on the full lattice and checked against finite differences (agreement to seven digits). Table~\ref{tab:sens} lists them for the two operating points and for the DFT-parameterized dot of Sec.~\ref{sec:dft}. Uniform in-plane fields \(F_xx+F_yy\) do not couple at first order: a field only displaces a harmonic dot, and the static polarizability \(1/2V_x\) of a harmonic oscillator is independent of the mass, so in the continuum limit both logical states shift identically to all orders. On the lattice the residual second-order shift is \(\lesssim50\)~peV for \(F=0.02\)~meV/nm. The leading electrostatic coupling is to the dot shape, through \(\delta\) and \(\bar V\). It is longitudinal with weight \(\cos\theta\) and vanishes at \(\delta=0\) by symmetry, since \(E_q(\delta)=E_q(-\delta)\), while \(f_R\) is maximal there [Fig.~\ref{fig:noise}(a,b)]. In real gate potentials, anharmonicity converts part of the dipolar field noise into curvature noise; this effect is not included here.

\begin{table}[!htbp]
\caption{Static sensitivities of the qubit frequency (\(L=40\)~nm). \(\phi_{x,y}\): small tilts of \(\mathbf L\) toward \(\hat x\) (the field direction) and \(\hat y\). \(\kappa\): fluctuator-ensemble conversion factor; \(T_2^*\) for \(\sigma_\mu=1~\mu\)eV.}
\label{tab:sens}
\begin{ruledtabular}
\begin{tabular}{lccc}
& Fig.~2 point & Fig.~3 point & DFT dot (Sec.~\ref{sec:dft})\\
\hline
\(f_q\) (GHz) / \(|\langle\sigma_z\rangle_{0,1}|\) & 12.51 / 0.749 & 17.20 / 0.687 & 18.52 / 0.815\\
\(|Q_{11}-Q_{00}|/2|Q_{01}|=\cot\theta\) & 1.13 & 0.95 & --\\
\(\partial f_q/\partial\delta\) (MHz per 0.01) & 239 & 208 & 4106\\
\(\partial f_q/\partial\ln\bar V\) (MHz per \(1\%\)) & 35 & 41 & 66\\
\(\partial f_q/\partial F_{x}\) (Hz per meV/nm) & \(0.1\) & \(0.1\) & --\\
\(\partial f_q/\partial B_x\), \(\partial f_q/\partial B_z\) (GHz/T) & 18.3, \(-21.0\) & 20.2, \(-19.2\) & 13.8, \(-22.8\)\\
\(\partial f_q/\partial B_y\), \(\partial f_q/\partial\phi_y\) & 0 & 0 & --\\
\(\partial f_q/\partial\ln J\) (MHz per \(1\%\)) & 67 & 78 & 91\\
\(\partial f_q/\partial\phi_x\) (MHz/mrad) & \(-6.3\) & \(-8.7\) & --\\
\(\kappa\) & \(9.2\times10^{-4}\) & \(3.6\times10^{-4}\) & \(1.6\times10^{-2}\)\\
\(T_2^*\) (\(\mu\)s), \(\sigma_\mu=1~\mu\)eV & 1.0 & 2.6 & 0.06\\
\end{tabular}
\end{ruledtabular}
\end{table}

\textit{Fluctuator model.}---Charge noise in gate-defined dots is commonly characterized by the fluctuations of the dot electrochemical potential, with spectral densities \(S_\mu^{1/2}(1~{\rm Hz})\approx0.3\)--\(1~\mu{\rm eV}/\sqrt{\rm Hz}\) in Si/SiGe and related platforms~\cite{PaqueletWuetz2023}. To convert such noise into qubit-frequency noise we place point-charge fluctuators at height \(h=10\)~nm above the layer, uniformly within a disk of radius 150~nm, with a screened Coulomb potential \(\propto e^{-d/\ell_s}/d\) (\(\ell_s=40\)~nm). We evaluate the exact lattice potential of each fluctuator without a multipole expansion. For 3000 samples we compute the first-order shifts of the electrochemical potential, \(\delta\mu=\langle0|\delta V|0\rangle\), and of the qubit energy, \(\delta E_q=\langle1|\delta V|1\rangle-\langle0|\delta V|0\rangle\). The ratio \(\kappa=\delta E_q^{\rm rms}/\delta\mu^{\rm rms}\) is independent of the fluctuator density and of \(\epsilon_r\). For quasi-static Gaussian noise, \(T_2^*=\sqrt2\hbar/(\kappa\sigma_\mu)\). For \(1/f\) noise integrated over \(10^{-2}\)--\(10^6\)~Hz, \(\sigma_\mu^2=S_\mu(1~{\rm Hz})\times(1~{\rm Hz})\ln10^8\), so \(\sigma_\mu\approx4.3\,S_\mu^{1/2}(1~{\rm Hz})\approx1.3\)--\(4.3~\mu\)eV, which gives \(T_2^*\approx0.6\)--\(2~\mu\)s at the Fig.~3 point. Dynamical decoupling would extend coherence beyond \(T_2^*\), but it is not simulated here. Figure~\ref{fig:noise}(d) shows that \(\kappa\) peaks at a lateral fluctuator distance \(R\approx15\)~nm and decreases beyond it: the qubit couples to the curvature of the fluctuating potential, which decays faster than the potential itself.

\textit{Coherence-time estimators.}---In each quasi-static noise realization the qubit energy shifts by \(\delta E_q\). The Ramsey coherence is the ensemble average \(|\langle e^{-i\delta E_qt/\hbar}\rangle|\), and \(T_2^*\) is its \(1/e\) time. For Gaussian \(\delta E_q\) with rms \(\sigma_E\) the coherence is \(e^{-\sigma_E^2t^2/2\hbar^2}\), so \(T_2^*=\sqrt2\hbar/\sigma_E\). This gives the first-order charge-noise expression above, with \(\sigma_E=\kappa\sigma_\mu\). It also gives the field-noise \(T_2^*\) quoted below: for a static field with rms \(\sigma_B\) per Cartesian component, representing field or Overhauser-field noise, \(\sigma_E=|\nabla_{\mathbf B}E_q|\,\sigma_B\), with the field sensitivities of Table~\ref{tab:sens}. Near \(\delta=0\) the first-order shape coupling vanishes, and second-order shifts become comparable to the small first-order remainder of the fluctuator potential. There we draw 4000 realizations of 40 fluctuators each from the model above. We add each potential to the Hamiltonian projected onto its lowest 24 eigenstates (orthonormalized also inside the degenerate shells at \(\delta=0\)), diagonalize exactly, and average the coherence directly. The potential is scaled so that the rms electrochemical-potential shift \(\langle0|\delta V|0\rangle\) equals \(\sigma_\mu\). This scale is fixed at the gate operating point in Sec.~\ref{sec:ensemble} and at the circular dot elsewhere; the rms shift changes by less than \(0.1\%\) between \(\delta=0\) and \(0.41\). Thermal excited-state populations are \([1+e^{E_q/k_BT}]^{-1}\). The echo time \(T_2\) is not computed; relaxation bounds it by \(T_2\le2T_1\).

\textit{Circular limit.}---The quadrupolar drive does not require a static ellipticity. At \(\delta=0\) the dot has the \([C_2\|C_4]\) symmetry of the \(d\)-wave altermagnet, a spin flip combined with a \(90^\circ\) rotation, which maps the spin-up ground orbital onto the spin-down one rotated by \(90^\circ\). At \(B=0\) the two are therefore degenerate (\(\Delta_{\rm AM}=0\)) but have opposite quadrupole moments \(Q_\sigma=\langle x^2-y^2\rangle_\sigma\). For the Fig.~3 set at the same \(\bar V\), \(Q_\uparrow=-Q_\downarrow=0.191~{\rm nm^2}\), and the splitting vanishes to numerical precision. The in-plane field mixes the two states through \(g_\perp\) into logical states with equal spin-up and spin-down weight (\(\langle\sigma_z\rangle=0\), \(\theta=\pi/2\)). The RF modulation of the ellipticity is then purely transverse, with the largest matrix element, \(|\langle1|Q|0\rangle|=|Q_\uparrow-Q_\downarrow|/2\). At \(B_x=0.45\)~T the circular dot has \(f_q=12.5\)~GHz, set by \(g_\perp\mu_BB_x\) alone, \(|\langle1|Q|0\rangle|=0.191~{\rm nm^2}\), \(28\%\) larger than at the Fig.~3 point, and \(E_2-E_0=9.6\)~meV. With the pulses of Table~\ref{tab:numerics} (8 and 16 cycles, \(\Delta t=0.5\)~ps, drive starting and ending at a node), \(A_Q=8.5\times10^{-6}~{\rm eV\,nm^{-2}}\) (\(2.8\%\) of \(\bar V\), 391~MHz Rabi frequency) gives \(\bar F=0.99982\) for \(X_{\pi/2}\) (0.64~ns) and \(0.99963\) for \(X_\pi\) (1.28~ns), with leakage \(\approx10^{-8}\). The Fig.~3 point gives 0.99966 and 0.99931.

The first-order sensitivities to \(\delta\), \(B_z\), and the N\'eel tilt \(\phi_x\) vanish. Those to \(\bar V\) and \(J\) drop from 41 and 78~MHz per \(1\%\) to 4~kHz and 2.0~MHz per \(1\%\), because both parameters now enter only through \(g_\perp\). This is the first-order sweet spot marked in Fig.~\ref{fig:noise}(b), where the shape contribution to the first-order \(T_2^*\) diverges. The rest of the fluctuator potential still shifts the two logical states slightly differently at first order, in proportion to \(B_x\); alone this would give \(T_2^*=0.48\)~s for \(\sigma_\mu=1~\mu\)eV and 0.12~s for \(4~\mu\)eV. With the exact-diagonalization estimator above, the fluctuator ensemble gives a charge-noise \(T_2^*\) of 0.40~s for \(\sigma_\mu=1~\mu\)eV, 65~ms for \(4~\mu\)eV, and 0.06--0.3~s over the range \(\sigma_\mu=1.3\)--\(4.3~\mu\)eV, compared with \(2.6~\mu\)s for \(1~\mu\)eV at the Fig.~3 point (Table~\ref{tab:sens}). Coherence is then limited by other channels. Field noise is not reduced: \(\partial f_q/\partial B_x=27.8\)~GHz/T replaces the combined \(B_x\) and \(B_z\) sensitivities of the elliptical point, so \(T_2^*\approx8~\mu\)s for \(\sigma_B=1~\mu\)T at both points. Hyperfine coupling and relaxation remain as well.

Static ellipticity is therefore not needed for the drive. It provides gate control of \(f_q\) at fixed \(B_x\) and the spin-sublattice polarization of the logical states that defines the operating contour of main-text Fig.~2(d). The circular limit trades both for the sweet spot. Its logical states are Zeeman-like, and perturbations that act through \(\Delta_{\rm AM}\) are maximally transverse there, so quadrupolar fields at \(\omega_q\) and shear strain relax the qubit rather than dephase it. The minimal model does not specify a strain response; for the DFT-parameterized material, this phonon relaxation is quantified in Sec.~\ref{sec:dft}, which also estimates the charge-noise relaxation.

\section{Quasi-static noise ensembles for the gates}
\label{sec:ensemble}

\begin{figure}[!htbp]
\centering
\includegraphics[width=0.96\textwidth]{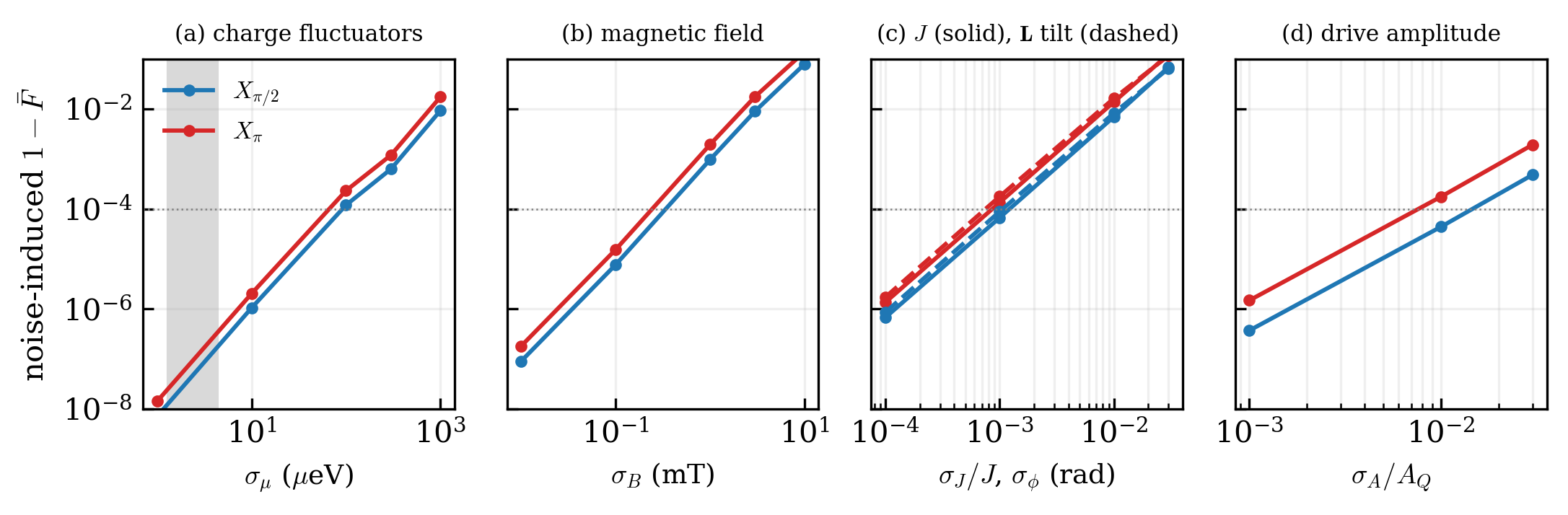}
\caption{Noise-induced average gate infidelity of the main-text \(X_{\pi/2}\) (blue) and \(X_\pi\) (red) gates under quasi-static noise (200 realizations per point): (a) charge fluctuators, with the range \(\sigma_\mu=1.3\)--\(4.3~\mu\)eV expected for \(S_\mu^{1/2}(1~{\rm Hz})=0.3\)--\(1~\mu{\rm eV}/\sqrt{\rm Hz}\) shaded; (b) static magnetic field; (c) exchange magnitude (solid) and N\'eel-vector tilt (dashed); (d) drive amplitude. The dotted line marks \(10^{-4}\).}
\label{fig:ensemble}
\end{figure}

To assess the gates under noise, we recompute the full driven evolution of main-text Fig.~3 (\(X_{\pi/2}\): 8 cycles, \(X_\pi\): 16 cycles, same calibrated amplitude and carrier phase) for 200 static noise realizations per amplitude. Each realization adds a perturbation \(\delta H\) projected onto the 64-state basis. The channels are: (a) charge noise from 40 random fluctuators per realization (model of Sec.~\ref{sec:noise}), scaled to an rms electrochemical-potential shift \(\sigma_\mu\); (b) an isotropic static field \(\delta\mathbf B\) with rms \(\sigma_B\) per component, representing field or Overhauser-field noise; (c) relative fluctuations \(\sigma_J/J\) of the staggered exchange and static tilts \(\phi_{x,y}\) of the N\'eel vector with rms \(\sigma_\phi\), representing order-parameter fluctuations; and (d) relative drive-amplitude errors \(\sigma_A/A_Q\). Figure~\ref{fig:ensemble} shows the noise-induced infidelity \(1-\bar F(U_0,U)\) relative to the noise-free gate \(U_0\), from which the leakage floor \(\approx10^{-6}\) of \(U_0\) is subtracted. The infidelity scales as \(\sigma^2\). For \(X_\pi\) it reaches \(10^{-4}\) at \(\sigma_\mu\approx60~\mu\)eV, \(\sigma_B\approx0.2\)~mT, \(\sigma_J/J\approx0.9\times10^{-3}\), \(\sigma_\phi\approx0.75\)~mrad, and \(\sigma_A/A_Q\approx0.75\%\). The realistic charge-noise range [shaded in Fig.~\ref{fig:ensemble}(a)] gives \(<10^{-6}\). Because the gates are short compared with \(T_2^*\), coherence rather than gate error limits operation (\(T_2^*/t_{\rm gate}\sim10^3\)). At millikelvin temperatures, thermal magnons are suppressed by the anisotropy gap, so the order-parameter fluctuations in (c) are expected to be far below the plotted range; the corresponding electron--magnon coupling is not modeled. Relaxation (\(T_1\)) is not part of these quasi-static ensembles. The minimal model does not specify how its hoppings respond to strain, so the phonon \(T_1\) of the main-text dots is undetermined. For the DFT-parameterized material, phonon relaxation through the spin--strain coupling is estimated in Sec.~\ref{sec:dft}, together with charge-noise relaxation for both parameter sets.

\textit{Static disorder.}---Device-to-device variations were modeled by adding a Gaussian-correlated random on-site potential (correlation length 3~nm, 40 realizations per amplitude) to the Fig.~3 dot. For rms amplitudes of 0.1, 0.3, and 1~meV, the qubit frequency spreads by \(\pm0.07\), \(\pm0.23\), and \(\pm1.0\)~GHz around 17.2~GHz, and the quadrupolar matrix element by \(\pm0.5\%\), \(\pm1.7\%\), and \(\pm7\%\). For 1~meV, the polarization is \(0.69\pm0.05\) (minimum 0.58) and the leakage gap \(7.4\pm0.4\)~meV. Every realization retains an opposite-polarized doublet, and \(95\%\) retain \(|\langle\sigma_z\rangle|\ge0.6\). Static disorder thus shifts the operating point, which must be calibrated per dot, but does not remove the doublet or the drive.

\section{DFT-parameterized dot}
\label{sec:dft}

\textit{Input.}---No published first-principles study that we are aware of reports spin-resolved band-edge masses at \(\Gamma\) for a two-dimensional \(d\)-wave altermagnetic semiconductor. The M\(_2\)X\(_2\)O monolayers have band edges at X and Y with spin--valley locking~\cite{Wu2024Fe2Se2O,Zou2024M2X2O}. The closest available input is the \(\Gamma\)-point \(k\cdot p\) fit of the two lowest conduction bands of the layered \(d\)-wave altermagnetic semiconductor La\(_2\)O\(_3\)Mn\(_2\)Se\(_2\) by Zhai \textit{et al.}~\cite{Zhai2025Strain}: \(H(\mathbf k,\eta)=\varepsilon_0+\rho_x(k_x^2+k_y^2)+\rho_zk_z^2+\lambda_{xy}k_xk_y\sigma_\chi+\tilde\lambda_{xy}\eta_{xy}\sigma_\chi\), with \(\rho_x=8.975~\mathrm{eV}\,\text{\AA}^2\), \(\lambda_{xy}=-9.742~\mathrm{eV}\,\text{\AA}^2\), and \(\tilde\lambda_{xy}=2.559\)~eV, fitted at 2\% shear strain. In axes rotated by \(45^\circ\), \(k_xk_y=(k_1^2-k_2^2)/2\), so Eq.~(\ref{eq:supp_band}) applies with \(A=\rho_x=89.75~{\rm meV\,nm^2}\) and \(\lambda=|\lambda_{xy}|/2=48.71~{\rm meV\,nm^2}\). The mean band mass is \(0.42\,m_e\), the spin-resolved masses are \(0.28\) and \(0.93\,m_e\), and \(\lambda/A=0.54\), compared with \(0.02\)--\(0.04\) for the main-text sets. The elliptical gate is then oriented along the crystal diagonals.

\textit{Lattice model.}---The \(k\cdot p\) fit determines \(A\) and \(\lambda\) but not the sublattice overlap \(\eta_{\rm sub}\), which sets \(g_\perp\). We choose \(\eta_{\rm sub}=0.90\) and \(\Delta_{ab}=25\)~meV, which with \(a=0.5\)~nm gives \(R_0=111.1\)~meV, \(J=96.9\)~meV, \(t_{AM}=111.7\)~meV, and \(t_1=-168.2\)~meV. This reproduces \(A\) and \(\lambda\) exactly, and the band minima of both spins remain at \(\Gamma\) over the full Brillouin zone. Larger \(\eta_{\rm sub}\) requires \(t_{AM}\) to approach \(|t_1|\), which moves the band minimum away from \(\Gamma\) (already for \(\eta_{\rm sub}=0.95\)).

\begin{figure}[!htbp]
\centering
\includegraphics[width=0.9\textwidth]{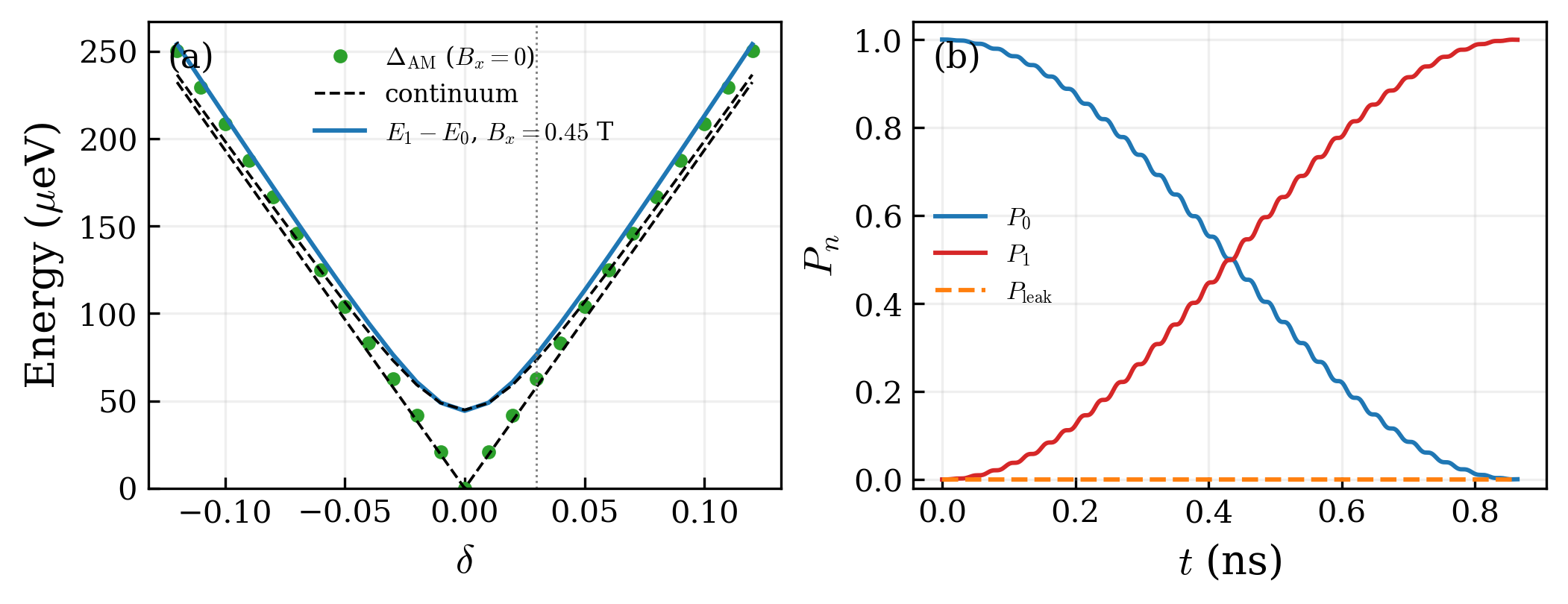}
\caption{DFT-parameterized dot (\(A\) and \(\lambda\) of La\(_2\)O\(_3\)Mn\(_2\)Se\(_2\)~\cite{Zhai2025Strain}, \(\eta_{\rm sub}=0.9\), \(\bar V=0.13~{\rm meV\,nm^{-2}}\), \(L=40\)~nm). (a) Field-free splitting \(\Delta_{\rm AM}\) (circles) and \(E_1-E_0\) at \(B_x=0.45\)~T (solid) versus \(\delta\); dashed lines are the continuum expressions, and the dotted line marks the operating point \(\delta=0.03\). (b) \(X_\pi\) pulse (16 cycles) at the operating point, \(A_Q=5.2\times10^{-7}~{\rm eV\,nm^{-2}}\).}
\label{fig:dft}
\end{figure}

\textit{Results} (Fig.~\ref{fig:dft}).---For \(\bar V=0.13~{\rm meV\,nm^{-2}}\) the atomistic splitting exceeds Eq.~(\ref{eq:supp_dam}) by \(\approx8\%\) because of band nonparabolicity, and reaches \(250~\mu\)eV at \(\delta=0.12\). Microwave-range qubits therefore require nearly circular dots. At \(\delta=0.03\) and \(B_x=0.45\)~T, \(f_q=18.5\)~GHz, \(\langle\sigma_z\rangle=\pm0.82\), \(E_2-E_0=4.25\)~meV, and \(|\langle1|Q|0\rangle|=4.6~{\rm nm^2}\), 31 times the Fig.~3 value. The calibrated 8- and 16-cycle pulses need only \(A_Q=5.2\times10^{-7}~{\rm eV\,nm^{-2}}\) (\(0.4\%\) of \(\bar V\)) for a 579~MHz Rabi frequency and give \(\bar F=0.99982\) (\(X_{\pi/2}\)) and \(0.99963\) (\(X_\pi\)) with leakage \(5\times10^{-7}\). The shape sensitivity is correspondingly larger (Table~\ref{tab:sens}): \(T_2^*=58\)~ns for \(\sigma_\mu=1~\mu\)eV at \(\delta=0.03\), increasing to \(0.21~\mu\)s at \(\delta=0.005\), \(B_x=0.45\)~T and to \(0.45~\mu\)s at \(\delta=0.005\), \(B_x=1\)~T, where \(|\langle\sigma_z\rangle|=0.11\). Strong anisotropy thus speeds up the drive at fixed RF amplitude and shifts the optimal operating point toward the circular sweet spot. At these high-field points, however, phonon relaxation through the strain coupling discussed next limits \(T_1\) to \(\approx0.1~\mu\)s at \(\delta=0.03\), \(B_x=0.45\)~T and to \(\approx16\)~ns at \(\delta=0.005\), \(B_x=1\)~T.

\begin{figure}[!htbp]
\centering
\includegraphics[width=0.9\textwidth]{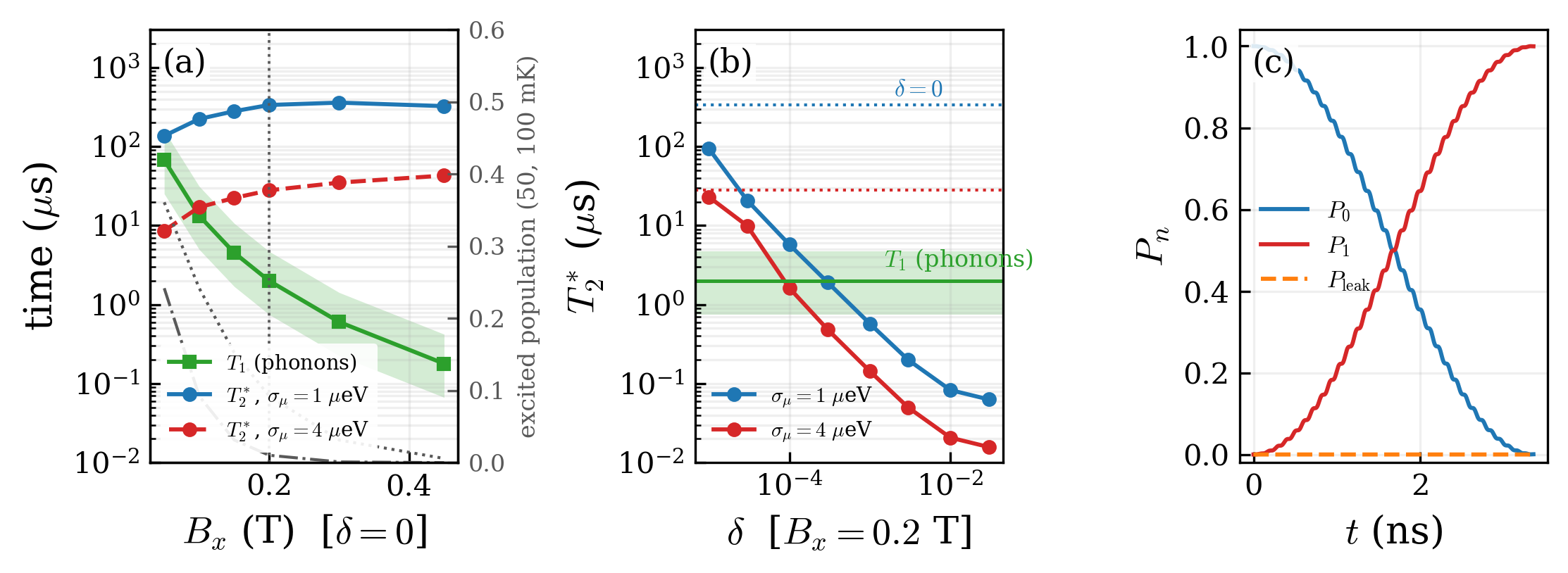}
\caption{Sweet-spot operation of the DFT-parameterized dot (\(\bar V=0.13~{\rm meV\,nm^{-2}}\), \(L=40\)~nm). (a) Field dependence at \(\delta=0\). Squares: phonon \(T_1\) at 50~mK from Eq.~(\ref{eq:phonon_t1}) for \(\rho=6~{\rm g\,cm^{-3}}\), \(v_t=3\)~km/s, \(v_l=5\)~km/s, with the band covering the phonon-parameter range. Circles: charge-noise-limited \(T_2^*\) for \(\sigma_\mu=1\) and \(4~\mu\)eV. Gray lines (right axis): thermal excited-state population at 50~mK (dash-dotted) and 100~mK (dotted). The vertical dotted line marks the chosen \(B_x=0.2\)~T. (b) \(T_2^*\) versus ellipticity at \(B_x=0.2\)~T, from exact shifts in the projected basis; horizontal dotted lines give the values at \(\delta=0\). The green line and band are the phonon \(T_1\). (c) Calibrated \(X_\pi\) pulse at \(\delta=0\) and \(B_x=0.2\)~T (16 cycles, 3.4~ns, \(A_Q=7.7\times10^{-8}~{\rm eV\,nm^{-2}}\)), starting at a drive node, with \(\bar F=0.9996\).}
\label{fig:dft_sweet}
\end{figure}

\textit{Strain and phonon relaxation.}---The same fit contains a shear-strain term \(\tilde\lambda_{xy}\eta_{xy}\sigma_\chi\) that splits \(\Gamma\) by \(2\tilde\lambda_{xy}\eta_{xy}\), equivalent to a field of 44~mT along \(\mathbf L\) per \(10^{-6}\) strain. Static strain from gate electrodes must therefore be compensated with the gate-tunable \(\Delta_{\rm AM}\), which is possible as long as \(|2\tilde\lambda_{xy}\eta_{xy}|\) does not exceed the attainable \(|\Delta_{\rm AM}|\). The same coupling makes acoustic phonons a direct spin-flip channel, without the spin-orbit suppression that protects conventional spin qubits. For isotropic acoustic phonons with density \(\rho\) and sound velocities \(v_l\), \(v_t\), one-phonon processes give
\begin{equation}
\frac{1}{T_1}=|\langle1|\sigma_{\mathbf L}|0\rangle|^2\,\frac{\tilde\lambda_{xy}^2\,\omega_q^3}{2\pi\rho\hbar}\left(\frac{1}{15v_l^5}+\frac{1}{10v_t^5}\right)\coth\frac{\hbar\omega_q}{2k_BT}.
\label{eq:phonon_t1}
\end{equation}
This follows from Fermi's golden rule for the coupling \(\tilde\lambda_{xy}\eta_{xy}\sigma_\chi\), where \(\sigma_\chi\equiv\sigma_{\mathbf L}\) is the spin along \(\mathbf L\). Here \(\eta_{xy}=(\partial_xu_y+\partial_yu_x)/2\) is a component of the symmetric strain tensor, as defined in Ref.~\cite{Zhai2025Strain}. The displacement field is that of isotropic bulk acoustic phonons in a volume \(\mathcal V\), \(\mathbf u=\sum_{\mathbf qs}[\hbar/(2\rho\mathcal V\omega_{\mathbf qs})]^{1/2}\hat{\mathbf e}_{\mathbf qs}(a_{\mathbf qs}e^{i\mathbf q\cdot\mathbf r}+{\rm h.c.})\). The phonon wavelength at \(\omega_q\), \(2\pi v_t/\omega_q\approx0.6~\mu\)m at 4.8~GHz, greatly exceeds the dot size, so the strain is uniform across the dot. The factors \(1/15\) and \(1/10\) are the angular averages of \(|\hat n_x\hat e_y+\hat n_y\hat e_x|^2/4\), with \(\hat{\mathbf n}=\mathbf q/q\), for longitudinal phonons and for the two transverse polarizations combined. Emission and absorption rates are proportional to \(n+1\) and \(n\), with \(n\) the Bose occupation at \(\omega_q\); their sum gives \(2n+1=\coth(\hbar\omega_q/2k_BT)\). Finally, \(|\langle1|\sigma_{\mathbf L}|0\rangle|^2=\sin^2\theta\) in the two-level description. Isotropic bulk phonons are an approximation for a layered compound and for a thin film on a substrate. We use \(\rho=6~{\rm g\,cm^{-3}}\), \(v_t=3\)~km/s, and \(v_l=5\)~km/s. The ranges \(\rho=5.5\)--\(6.5~{\rm g\,cm^{-3}}\), \(v_t=2.5\)--\(3.5\)~km/s, and \(v_l=4.5\)--\(6\)~km/s change \(T_1\) by a factor of 2--3 either way. At the elliptical point above (\(f_q=18.5\)~GHz) this gives \(T_1\approx0.1~\mu\)s. Because \(T_1^{-1}\propto\sin^2\theta\,f_q^3=(g_\perp\mu_BB_x/h)^2f_q\), \(T_1\) is longest at the circular point for a given field and grows rapidly as the field is lowered [Fig.~\ref{fig:dft_sweet}(a)]. Two other phonon channels are negligible. Scalar deformation-potential coupling acts only through the quadrupole moment of the dot and is suppressed by \((q^2|\langle1|Q|0\rangle|)^2\sim10^{-6}\) at these phonon wavelengths. Single-magnon emission is forbidden when the magnon gap exceeds \(\hbar\omega_q\).

\textit{Sweet-spot operation at \(B_x=0.2\)~T.}---The longitudinal shape coupling carries the weight \(\cos\theta=\Delta_{\rm AM}/(E_1-E_0)\), so the strongly anisotropic dot is best operated at the circular point, where \(\Delta_{\rm AM}\) vanishes by symmetry (Sec.~\ref{sec:noise} gives this circular limit for the Fig.~3 set). There:
\begin{itemize}
\item the logical states are quantized along the field (\(\langle\sigma_z\rangle=0\)), and the quadrupolar drive is purely transverse with its largest matrix element (\(|\langle1|Q|0\rangle|=8.0~{\rm nm^2}\));
\item every perturbation that acts through \(\Delta_{\rm AM}\) alone (dot shape, shear strain, and N\'eel-vector tilt) enters the qubit energy only at second order, so dephasing from these sources is weak;
\item the exchange magnitude also sets \(g_\perp\) and therefore stays first order: \(\partial f_q/\partial\ln J=-14.6\)~MHz per \(1\%\) at 0.2~T, i.e., \(T_2^*\approx0.15~\mu{\rm s}\times10^{-3}/(\sigma_J/J)\), so that coherence near the \(T_1\) limit requires \(\sigma_J/J\lesssim4\times10^{-5}\) (thermal order-parameter fluctuations should be far weaker at millikelvin temperatures, Sec.~\ref{sec:ensemble});
\item strain and quadrupolar fields at \(\omega_q\) are, however, maximally transverse, so relaxation through Eq.~(\ref{eq:phonon_t1}) is not suppressed;
\item field noise along \(B_x\) remains first order (\(\partial f_q/\partial B_x=23.8\)~GHz/T, i.e., \(T_2^*=9.5~\mu\)s for \(\sigma_B=1~\mu\)T).
\end{itemize}
To include the second-order charge-noise contribution, we use the exact-diagonalization estimator of Sec.~\ref{sec:noise}. At \(B_x=0.2\)~T and \(\delta=0.03\) it gives 63~ns at \(\sigma_\mu=1~\mu\)eV, equal to its first-order value, so the second-order contribution matters only near \(\delta=0\).

The field is a compromise [Fig.~\ref{fig:dft_sweet}(a), \(\delta=0\)]. Lowering \(B_x\) lengthens the phonon \(T_1\) rapidly, roughly as \(B_x^{-3}\): at 50~mK it is 0.18, 0.60, 2.0, 4.5, 13, and 67~\(\mu\)s for \(B_x=0.45\), 0.3, 0.2, 0.15, 0.1, and 0.05~T. Lowering the field also has three costs:
\begin{itemize}
\item the thermal excited-state population grows: at 50 (100)~mK it is 1\% (9\%) at 0.2~T, 9\% (24\%) at 0.1~T, and 24\% (36\%) at 0.05~T, which compromises initialization by thermalization and energy-selective readout;
\item the second-order charge-noise \(T_2^*\) at \(\delta=0\) falls (for \(\sigma_\mu=4~\mu\)eV: 43, 28, and \(17~\mu\)s at 0.45, 0.2, and 0.1~T), and the first-order slope near \(\delta=0\) scales as \(1/B_x\), which tightens the sweet-spot tolerance;
\item the gates slow down, because the Rabi frequency must remain well below \(f_q\).
\end{itemize}
We take \(B_x=0.2\)~T (\(f_q=4.76\)~GHz). There \(T_1=2.0~\mu\)s at 50~mK (0.75--\(4.7~\mu\)s across the phonon parameters; \(1.7~\mu\)s at 100~mK), and the thermal population is 1\% at 50~mK.

Figure~\ref{fig:dft_sweet}(b) shows the charge-noise-limited \(T_2^*\) at this field: 335~\(\mu\)s (\(\sigma_\mu=1~\mu\)eV) and \(28~\mu\)s (\(4~\mu\)eV) at \(\delta=0\), and 206 and \(24~\mu\)s for the range \(\sigma_\mu=1.3\)--\(4.3~\mu\)eV of Sec.~\ref{sec:noise}. It stays above \(2T_1\approx4~\mu\)s for \(|\delta|\lesssim10^{-4}\) (\(\sigma_\mu=1~\mu\)eV, \(T_2^*=5.7~\mu\)s) or \(|\delta|\lesssim3\times10^{-5}\) (\(4~\mu\)eV, \(T_2^*=9.9~\mu\)s), so charge noise leaves coherence \(T_1\)-limited, \(T_2\le2T_1\). The calibrated 8- and 16-cycle gates [Fig.~\ref{fig:dft_sweet}(c)] need only \(A_Q=7.7\times10^{-8}~{\rm eV\,nm^{-2}}\) (\(0.06\%\) of \(\bar V\)) for a 149~MHz Rabi frequency. Starting at a drive node, they give \(\bar F=0.99982\) for \(X_{\pi/2}\) (1.7~ns) and \(0.99963\) for \(X_\pi\) (3.4~ns), with leakage below \(10^{-8}\). An optimized start phase cancels the counter-rotating error and gives \(\bar F>0.9999\) with leakage below \(10^{-7}\) (Sec.~\ref{sec:convergence}). \(T_1\) corresponds to about 600 \(X_\pi\) durations.

Staying within \(|\delta|\lesssim10^{-4}\) amounts to locating the minimum of \(f_q(\delta)\), which is quadratic: at 0.2~T, \(f_q\) rises by \(\approx0.3\)~MHz at \(\delta=10^{-4}\) and by 27~MHz at \(10^{-3}\). The same tuning absorbs static offsets of the longitudinal splitting from strain or disorder, because the sweet spot is the point where the total splitting along \(\mathbf L\) vanishes. The price is the loss of spin-sublattice polarization of the logical states, which become Zeeman-like states driven electrically through the altermagnetic channel. For a material with a ten times weaker spin--strain coefficient, \(T_1\) would be a hundred times longer, allowing higher fields and simpler readout.

\textit{Charge-noise relaxation.}---The fluctuator potential is also maximally transverse at \(\delta=0\), so charge noise at \(\omega_q\) relaxes the qubit. For independent fluctuators with a common spectrum, the transverse matrix element \(X=\langle1|\delta V|0\rangle\) has the spectral density \(S_X(f)=\kappa_\perp^2S_\mu(f)\), where \(\kappa_\perp=X^{\rm rms}/\delta\mu^{\rm rms}=8.9\times10^{-3}\) for the ensemble of Sec.~\ref{sec:noise}, independent of \(B_x\) at \(\delta=0\). For classical noise, \(1/T_1=S_X(f_q)/\hbar^2\), with \(S\) the one-sided spectral density. Extrapolating \(S_\mu(f)=S_\mu(1~{\rm Hz})\,(1~{\rm Hz}/f)\) to \(f_q=4.76\)~GHz gives \(T_1=26~\mu\)s for \(S_\mu^{1/2}(1~{\rm Hz})=1~\mu{\rm eV}/\sqrt{\rm Hz}\) and \(290~\mu\)s for \(0.3~\mu{\rm eV}/\sqrt{\rm Hz}\), the range that gives \(\sigma_\mu=1.3\)--\(4.3~\mu\)eV. For \(1/f\) noise this \(T_1\) grows in proportion to \(f_q\), opposite to the phonon \(T_1\), which is the shorter of the two at 0.2~T. The \(1/f\) form is extrapolated far above the frequencies at which it is measured, so this is an order-of-magnitude estimate: a white or Johnson-noise floor at gigahertz frequencies would shorten \(T_1\), and a faster high-frequency decay would lengthen it. For the main-text Fig.~3 set, whose \(\kappa_\perp\) is 40--50 times smaller, the same estimate gives \(T_1=0.23\)--\(2.5\)~s at the gate point and \(0.11\)--\(1.2\)~s at its circular point.

\textit{Caveats.}---La\(_2\)O\(_3\)Mn\(_2\)Se\(_2\) is a bulk layered compound (we neglect \(\rho_zk_z^2\), appropriate for a thin film). The fit was performed under strain, and the global location of its conduction-band minimum is not established in Ref.~\cite{Zhai2025Strain}. Moreover, \(^{139}\)La (\(I=7/2\)) and \(^{55}\)Mn (\(I=5/2\)) are essentially 100\% abundant, so the compound is a test of the mechanism with first-principles coefficients rather than a qubit host candidate. A Wannier-based atomistic model of a monolayer candidate is the natural next step.

\section{Two-electron CI basis convergence}

\begin{figure}[!htbp]
\centering
\includegraphics[width=0.9\textwidth]{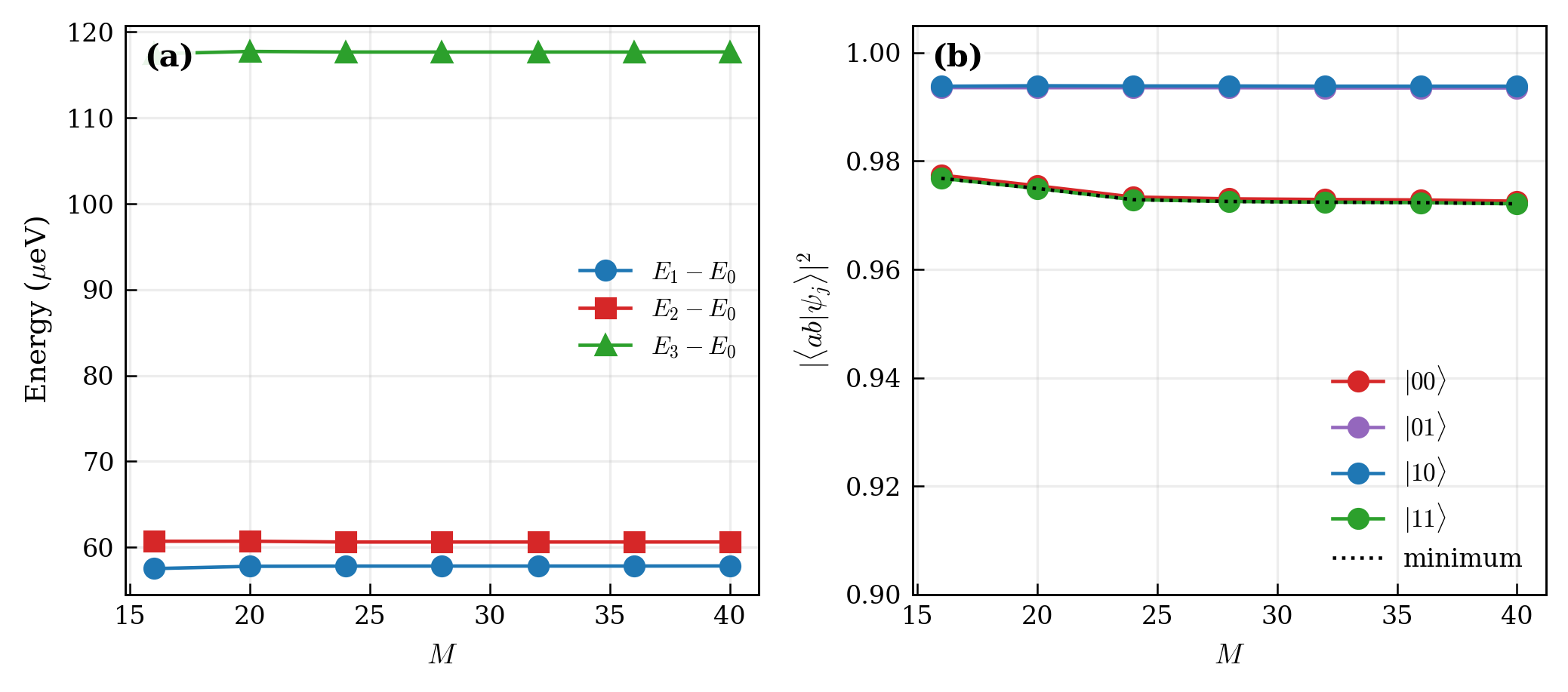}
\caption{Convergence of the two-electron CI calculation with the number \(M\) of retained one-electron orbitals at the main-text Fig.~4 operating point. (a) Four-state energy gaps \(E_1-E_0\), \(E_2-E_0\), and \(E_3-E_0\) versus \(M\). (b) Assignment-aware fixed isolated-dot product overlaps for the four lowest dressed CI states; the dotted line gives the minimum over the four assignments. Parameters: \(a=0.5\)~nm, \(L=60\)~nm, \(d=18.5\)~nm, \(\epsilon=4\)~meV, \(V_b^{(0)}=0\), left-dot curvatures \((V_x^L,V_y^L)=(0.185,0.434)~{\rm meV\,nm^{-2}}\), right-dot curvatures \((V_x^R,V_y^R)=(0.434,0.185)~{\rm meV\,nm^{-2}}\), \(V_s=0.2\)~meV, \((w_x,w_y)=(8,12)\)~nm, \(\epsilon_r=24\), \(d_{\rm soft}=2.5\)~nm, \(\ell_s=40\)~nm, \(J=12\)~meV, \(\Delta_{ab}=12\)~meV, \(B_x=0.3\)~T, no SOC, \(t_1=-150\)~meV, \(t_2=0\), \(t_{AM}=15\)~meV, \(\mu=0\).}
\label{fig:supp_twoqubit_mconv}
\end{figure}

The two-electron calculation uses a truncated CI basis built from the lowest \(M\) one-electron spinor eigenstates of the static double-dot Hamiltonian. Figure~\ref{fig:supp_twoqubit_mconv} shows the dependence of the low-energy CI spectrum and of the fixed isolated-dot product overlaps on \(M\) at the operating point of main-text Fig.~4. The assignment of the four lowest dressed CI states remains \(|10\rangle_{\rm loc}\), \(|00\rangle_{\rm loc}\), \(|11\rangle_{\rm loc}\), and \(|01\rangle_{\rm loc}\) over the plotted range, and the minimum assigned product overlap at \(M=32\) is \(0.972\).

\section{Full two-qubit process of the barrier pulse}
\label{sec:twoqubit}

The main-text Fig.~4(d) propagates only the dressed state assigned to \(|10\rangle_{\rm loc}\). Here the same pulse (32 dressed CI states, barrier amplitude 0.500~meV, carrier 28.45~GHz, 561 cycles, 19.72~ns, midpoint step 2~ps) is applied to all four dressed logical states, which gives the \(4\times4\) logical block \(M\) of the propagator. We refer \(M\) to two frames. In the dressed frame, each dressed state's own energy is removed. In single-qubit frames, only the local part of the dressed energies is removed, so the conditional phase \(\zeta t/\hbar\) from the static coupling \(\zeta=E_{00}+E_{11}-E_{01}-E_{10}=0.754~\mu\)eV is retained. Here \(\zeta\) is defined from the dressed CI energies. It is not \(4h_{zz}=-0.16~\mu\)eV of the fixed localized-basis decomposition in the main text, which is a first-order projection. Couplings inside that block barely change this value: mutual repulsion of \(|01\rangle\) and \(|10\rangle\) leaves \(E_{01}+E_{10}\) unchanged, and diagonalizing \(H_{\rm log}\) gives \(-0.15~\mu\)eV. The difference comes from virtual coupling to the non-logical CI states, whose second-order shift of \(\zeta\) is \(+0.96~\mu\)eV. The same states dominate the rest of the low-energy Hamiltonian. Block-diagonalizing the four dressed states in the localized basis gives a nonlocal block of norm \(1.0~\mu\)eV with \(h_{zz}=+0.20~\mu\)eV and \((h_{xz}^2+h_{yz}^2)^{1/2}=(h_{zx}^2+h_{zy}^2)^{1/2}=0.56~\mu\)eV, compared with 0.22, \(-0.04\), and \(0.12~\mu\)eV for \(H_{\rm log}\). The separate \(x\) and \(y\) components depend on the arbitrary phases of the reference states. In particular \(|\langle00|H|11\rangle|=0.40~\mu\)eV, so with \(E_2-E_1=2.80~\mu\)eV the localized \(|00\rangle\) and \(|11\rangle\) exchange up to \(8\%\) of their population at idle with a 1.5~ns period. The dressed states are therefore the natural logical basis. We characterize \(M\) by its leakage, by the Makhlin local invariants \(G_{1,2}\)~\cite{Makhlin2002} and entangling power \(e_p=\tfrac29(1-|G_1|)\), by the perfect-entangler criterion of Ref.~\cite{Zhang2003}, and by the best average gate fidelity to an fSim\((\theta,\varphi)\) gate~\cite{Foxen2020} dressed by single-qubit \(Z\) rotations before and after (virtual-\(Z\) frame updates). The results are summarized in Table~\ref{tab:twoqubit}. At half the pulse (\(t=9.86\)~ns; the Bell fidelity of main-text Fig.~4(d) peaks at 9.83~ns) the pulse realizes an fSim gate with \(\theta\simeq\pi/4\), a perfect entangler with the entangling power of \(\sqrt{\rm iSWAP}\) (\(e_p=1/6\)), with \(\bar F=0.99996\); without the conditional phase, i.e., in the dressed frame, it is locally equivalent to \(\sqrt{\rm iSWAP}\) (\(G_1=1/4\), \(G_2=1\)). At the full transfer, the static conditional phase makes the gate close to SWAP-like and no longer a perfect entangler in single-qubit frames, so the half-transfer pulse is the useful entangling operation. Halving the time step changes the fidelity and the leakage by \(\lesssim4\times10^{-6}\) and the fitted \(\theta\) by 0.003, because the midpoint rule slightly under-drives at 2~ps.

\begin{table}[!htbp]
\caption{Two-qubit process of the resonant barrier pulse (dressed CI basis; values for a 1~ps step in parentheses).}
\label{tab:twoqubit}
\begin{ruledtabular}
\begin{tabular}{lcc}
& half pulse, \(t=9.86\)~ns & full transfer, \(t=19.72\)~ns\\
\hline
logical leakage \(1-{\rm Tr}(M^\dagger M)/4\) & \(1.8\times10^{-5}\) (\(1.5\times10^{-5}\)) & \(8\times10^{-6}\) (\(1.2\times10^{-5}\))\\
dressed frame: \(G_1\), \(G_2\) & \(0.247-0.024i\), \(0.986\) & \(-0.009\), \(-1.015\)\\
single-qubit frames: \(G_1\), \(G_2\) & \(0.093+0.231i\), \(0.373\) & \(-0.852\), \(-2.704\)\\
single-qubit frames: best fSim \((\theta,\varphi)\) & \((0.788,\,5.10)\) (\((0.791,\,5.10)\)) & \((1.60,\,3.93)\)\\
\(\bar F\) to fSim with virtual \(Z\) & \(0.99996\) (\(0.99996\)) & \(0.99997\) (\(0.99997\))\\
entangling power \(e_p\) (\(\sqrt{\rm iSWAP}\): 1/6) & \(0.167\) (\(0.167\)) & \(0.033\)\\
perfect entangler (single-qubit frames) & yes & no\\
\end{tabular}
\end{ruledtabular}
\end{table}

\end{document}